\documentclass[12pt,preprint]{emulateapj}

\usepackage{natbib}
\usepackage{amsmath}
\usepackage{hyperref}
\usepackage{color}

\slugcomment{May 10, 2012}

\shorttitle{Evolution of Gravitationally Unstable Disks}
\shortauthors{Forbes, Krumholz, \& Burkert}

\begin{document}

\title{Evolving Gravitationally Unstable Disks Over Cosmic Time: Implications For Thick Disk Formation}

\author{John Forbes\altaffilmark{1}, Mark Krumholz\altaffilmark{2}}
\affil{Department of Astronomy \& Astrophysics, University of California, Santa Cruz, CA 95064 USA}

\author{Andreas Burkert\altaffilmark{3}}
\affil{University Observatory Munich (USM), Scheinerstrasse 1, 81679 Munich, Germany and Max-Planck-Institut fuer extraterrestrische Physik, Giessenbachstrasse 1, 85758 Garching, Germany, Max-Planck-Fellow}

\altaffiltext{1}{E-mail: jforbes@ucolick.org}
\altaffiltext{2}{E-mail: krumholz@ucolick.org}
\altaffiltext{3}{E-mail: burkert@usm.uni-muenchen.de}

\begin{abstract}
Observations of disk galaxies at $z\sim 2$ have demonstrated that turbulence driven by gravitational instability can dominate the energetics of the disk. We present a 1D simulation code, which we have made publicly available, that economically evolves these galaxies from $z\sim 2$ to $z\sim 0$ on a single CPU in a matter of minutes, tracking column density, metallicity, and velocity dispersions of gaseous and multiple stellar components. We include an H$_2$ regulated star formation law and the effects of stellar heating by transient spiral structure. We use this code to demonstrate a possible explanation for the existence of a thin and thick disk stellar population and the age-velocity dispersion correlation of stars in the solar neighborhood: the high velocity dispersion of gas in disks at $z\sim 2$ decreases along with the cosmological accretion rate, while at lower redshift, the dynamically colder gas forms the low velocity dispersion stars of the thin disk.
\end{abstract}


\section{Introduction}

In the past decade, observations of galaxies near $z \sim 2$ have revealed compelling evidence for the importance of gravitational instability in their dynamics and evolution. A whole class of galaxies has been observed whose images are dominated by large luminous clumps of gas \citep{Elmegreen:2004gs,Elmegreen:2005km,Forster-Schreiber:2009gd}, while measurements of the velocity dispersion of such massive star-forming galaxies have found values near 50 km/s spread across the entire disk \citep{Cresci:2009ta,Genzel:2011dd}. This is difficult to reproduce with supernova feedback, which is strongest near the centers of galaxies where the star formation rate peaks, and which is only strong enough to drive velocity dispersions of $\sim 10$ km/s \citep{Joung:2009ec}. Other forms of stellar feedback may drive turbulence \citep{Thompson:2005vy,Elmegreen:2010iw}, but we will concentrate on the case where turbulence is driven by gravitational instability in the disk.

To a first approximation, the gravitational stability of a thin disk to axisymmetric perturbations is described by Toomre's Q parameter $Q=\kappa \sigma/(\pi G \Sigma)$, where $\kappa$ is the epicyclic frequency, $\sigma$ is the 1d velocity dispersion, and $\Sigma$ is the gas surface density. The disk is unstable when $Q\lesssim 1$. The importance of gravitational instability in high redshift galaxies arises from the high cosmological accretion rates they experience, which drive up the value of $\Sigma$ \citep{Dekel:2009xh}. This instability gives rise to clumps of the sort observed at high redshift. The inhomogenous and time-varying gravitational field drives turbulence throughout the disk, regardless of the stellar density or supernova rate. The energy source for these random motions must ultimately be the gravitational potential of the galaxy, so gas is transported inwards.

Cosmological simulations with sufficiently high resolution \citep{Bournaud:2009jb, Ceverino:2010mq} successfully reproduce disks in which gravitational instability forms clumps and causes the inward migration of material through galactic disks. Simulations of isolated galaxies \citep{Bournaud:2011fb,2011MNRAS.417.1318D,2011MNRAS.413.2935D} with initial conditions set such that $Q < 1$ provide a higher resolution view of such galaxies over a few outer orbits of the disk. These studies, while illuminating, are expensive, since they must solve the equations of hydrodynamics in three dimensions over cosmological times. The model we present here solves the hydrodynamical equations in the limit of a thin axisymmetric disk. Since quantities vary only in the radial direction, the problem is computationally much cheaper to solve, allowing us to explore parameter space efficiently, while still solving the full 1D equations of fluid dynamics instead of relying entirely on semi-analytic models \citep{Dekel:2009xh,Cacciato:2011ab}
     
	Past 1D models of gravitational instability in disks have a number of shortcomings. The rate at which mass and angular momentum are transported inwards is often parameterized and fit to the results of hydrodynamical simulations, rather than being derived from first principles. The rotation curves are only allowed to be either Keplerian or flat. Energy is frequently assumed to be instantaneously equilibrated, which neglects the possibility that it might be advected through the disk. The pressure support of the disk is often treated as coming from thermal pressure rather than supersonic turbulence. Few models take into account the stellar component of the disk, which becomes increasingly important as the galaxy evolves towards the present day, and can ultimately provide a variety of observable predictions.
	
	In particular, the age-velocity dispersion-metallicity correlation of stars in the solar neighborhood \citep{Nordstrom:2004av}, might well be explained by means of gravitational instability in high redshift disks \citep{2009ApJ...707L...1B}. The high velocity dispersion in these disks means that the population of stars formed in that epoch will start with a high velocity dispersion \citep{Burkert:1992th}. The gas disk cools as a result of slowing cosmological accretion rates, so younger stars are formed in a thinner, more metal-rich disk. This mechanism of thick and thin disk formation contrasts with the more common story that various secular processes and minor mergers heat thin disk stars into a thick disk \citep[e.g.][]{1987gady.book.....B}. 
	
	\citet{Krumholz:2010kc} (hereafter KB10) found an analytic steady-state solution to the full equations of fluid dynamics in the thin disk limit under the assumption that the disk self-regulates to maintain $Q=1$. To make the problem tractable analytically, however, they required a handful of simplifying assumptions: they use an analytic approximation to Q, which becomes progressively worse at lower redshift as the ratio of gas to stellar velocity dispersion deviates from unity. They also assume that the velocity dispersion of stars and gas are equal, and the gas fraction at all points in the disk remains constant in radius and time. In this paper, we relax these assumptions and include treatments of stellar migration, metallicity, the non-zero thermal temperature of the gas, and evolution of individual stellar populations. These improvements along with an efficient simulation code allow us to realistically evolve disks from high redshift to the present day at minimal computational expense.

	In section 2 we derive the equations governing the evolution of the gas over time. Section 3 presents the derivation of the equations governing the stellar dynamics. In section 4, we derive the evolution of metallicity in the gas and stars. In section 5 we discuss how these differential equations are solved numerically, and in section 6 we present the results for fiducial parameters chosen to be similar to the Milky Way. We conclude in section 7. The code we describe, named GIDGET for Gravitational Instability-Dominated Galaxy Evolution Tool, is available at \url{http://www.ucolick.org/~jforbes/gidget.html}
	
\section{Gas Evolution Equations}
\subsection{Basic Equations}
We first give a brief overview of the derivation of the evolution equations for the gas column density $\Sigma$ and velocity dispersion $\sigma$. For more details see KB10. The equations of mass, momentum, and energy conservation for a viscous star-forming fluid in a gravitational field are
\begin{mathletters}
\begin{eqnarray}
\frac{\partial \rho}{\partial t} &=& -\nabla \cdot (\rho {\bf v}) - (f_R+\mu)\dot{\rho}_* , \\
\label{navierstokes}
\rho \frac{D{\bf v}}{D t} &=& -\nabla P - \rho \nabla \psi + \nabla \cdot {\bf T}, \\
\rho \frac{D e}{D t} &=& -P \nabla \cdot {\bf v} + \Phi + \Gamma - \Lambda, 
\end{eqnarray}
\end{mathletters}
where $\rho$, ${\bf v}$, $e$, and $P$ are the gas density, velocity, specific internal energy, and pressure respectively. The star-formation rate per unit volume at an Eulerian point is $\dot{\rho}_*$, with a mass loading factor $\mu$ equal to the ratio of gas ejected in \textcolor{black}{galactic-scale} winds to the star formation rate. We will be employing the instantaneous recycling approximation (see section \ref{sec:metallicity}), which approximates all stellar evolution as occurring immediately. Of the mass which forms stars, the gas will only lose the so-called remnant fraction, $f_R$, to stars, while the remaining $(1-f_R)$ will be immediately recycled into the ISM. The gravitational potential is $\psi$, ${\bf T}$ is the viscous stress tensor, $\Phi = T^{ik} (\partial v_i/\partial x_k)$ is the rate of viscous dissipation, and $\Gamma$ and $\Lambda$ are the rates of radiative energy gain and loss per unit volume.

For a thin disk, we formally have $\rho=\Sigma \delta(z)$ \textcolor{black}{and $v_z=0$}. By expanding the fluid equations in cylindrical coordinates, \textcolor{black}{integrating over z,} assuming axisymmetry and $v_r \ll v_\phi$, and dropping time derivatives of the potential and the circular velocity, we can obtain evolution equations for the gas column density and gas velocity dispersion. The evolution of column density is given by
\begin{eqnarray}
\frac{\partial\Sigma}{\partial t} &=& \frac{1}{2\pi r}\frac{\partial}{\partial r} \dot{M} - (f_R+\mu)\dot{\Sigma}_*^{SF}  \nonumber \\
\label{eq:dcoldt}
 &=&  \frac{1}{2 \pi (\beta + 1) r v_\phi}\left[ \frac{\beta(\beta+1) + r (\partial\beta/\partial r)}{(\beta+1) r} \left( \frac{\partial \mathcal{T}}{\partial r} \right) - \frac{\partial^2 \mathcal{T}}{\partial r^2} \right] \nonumber \\
 & &  - (f_R+\mu)\dot{\Sigma}_*^{SF}
\end{eqnarray}
where $\beta = \partial \ln v_\phi / \partial \ln r$ is the power law index of the rotation curve, $\mathcal{T} = \int 2\pi r^2 T_{r\phi} dz$ is the viscous torque, $\dot{\Sigma}_*^{SF}$ is the star formation rate per unit area, and
\begin{equation}
\label{mdot}
\dot{M} = -2\pi r\Sigma v_r = -\frac{1}{v_\phi(1+\beta)} \frac{\partial \mathcal{T}}{\partial r}
\end{equation}
is the mass flux. The second equality follows from the angular momentum equation, which is in turn derived from the $\phi$ component of the Navier-Stokes equation (equation \ref{navierstokes}\textcolor{black}{, see KB10}).

The derivation of the velocity dispersion evolution equation requires an equation of state, which we take to be $P=\rho \sigma^2$. The velocity dispersion has a thermal and a turbulent component. It is a reasonable approximation to treat both as contributing to the pressure so long as we average over scales much larger than the characteristic size of the turbulent eddies, which will be of order the disk scale height.

Taking the dot product of the velocity with the Navier-Stokes equation and adding it to the internal energy equation yields an equation for the total energy, i.e. internal energy, kinetic energy, and gravitational potential energy. By decomposing the velocity as $v^2 = v_r^2 + v_\phi^2 + 3\sigma_{nt}^2$, the kinetic plus thermal energy may be rewritten
\begin{equation}
\frac{1}{2} v^2 + e = \frac{1}{2}(v_r^2 + v_\phi^2) + \frac{3}{2} \sigma^2
\end{equation}
where the velocity dispersion is taken to be the quadrature sum of a thermal and non-thermal component, $\sigma^2 = \sigma_t^2+ \sigma_{nt}^2$. Neglecting the $v_r^2$ term as small compared to both $\sigma^2$ and $v_\phi^2$ in a thin, rotation-dominated, $Q\sim 1$ disk, employing radial force balance to set $\partial \psi/\partial r = v_\phi^2/r$, assuming a constant potential to set $\partial v_\phi / \partial t = 0$, and integrating over $z$ yields the evolution equation
\begin{eqnarray}
\label{eq:dsigdt}
\frac{\partial \sigma}{\partial t} &= &\frac{\mathcal{G}-\mathcal{L}}{3\sigma \Sigma} + \frac{1}{6\pi r \Sigma} \Bigg[ (\beta-1)\frac{v_\phi}{r^2\sigma} \mathcal{T}  \nonumber \\ 
&  & + \frac{\beta^2\sigma + \sigma (r \frac{d\beta}{d r}+\beta) - 5(\beta+1)r\frac{\partial \sigma}{\partial r}}{(\beta+1)^2 r v_\phi} \left(\frac{\partial \mathcal{T}}{\partial r}  \right)  \nonumber \\
&  & - \frac{\sigma}{(\beta+1)v_\phi}\left(\frac{\partial^2 \mathcal{T}}{\partial r^2}  \right) \Bigg]  
\end{eqnarray}
To fully specify the evolution of the gas, we need to set a rotation curve, a prescription for radiative energy gain and loss per unit area, and a procedure for finding the viscous torque. This will allow us to specify $v_\phi$, $\beta$, $\mathcal{G}=\int\Gamma dz$, $\mathcal{L}=\int\Lambda dz$, and $\mathcal{T}$. The rotation curve is specified at run-time, and $\mathcal{T}$ is set by our treatment of  gravitational instability (see section \ref{torque}). We set $\mathcal{G}=0$, which is equivalent to requiring that the energy balance in the gas is completely determined by the effects of the viscous torque and radiative loss. We assume that the loss rate, meanwhile, is proportional to the kinetic energy density per disk scale height crossing time, in agreement with the decay rate of turbulence observed in full 3D MHD simulations of supersonic turbulence \citep{Mac-Low:1998mk, Stone:1998on}.
\begin{equation}
\mathcal{L} \equiv \frac{d}{dt}\left(\frac{3}{2}\Sigma\sigma^2\right)^{rad} = \frac{d}{dt}\left(\frac{3}{2}\Sigma\sigma_{nt}^2\right)^{rad}  =   \eta \frac{\Sigma\sigma_{nt}^2}{H/\sigma_{nt}}
\end{equation}
where $\eta$ is a free parameter of order unity.  If the decay time is exactly the crossing time, $\eta=1.5$, since the kinetic energy surface density is $(3/2)\Sigma\sigma^2$. In dropping the time derivative of $\sigma_t$, we have assumed that the thermal velocity dispersion is unaffected by radiative dissipation, i.e. that the gas is isothermal. 

\textcolor{black}{The scale height $H$ is approximated by taking the solution to the equations of vertical equilibrium for a single component disk, $H_1=\sigma^2/\pi G\Sigma$, and adopting it to multiple components: 
\begin{equation}
\label{eq:scaleheight}
H = \frac{\sigma^2}{\pi G (\Sigma + f \Sigma_* )},
\end{equation}
where $f$ represents the relative importance of the stellar mass, or the stellar mass within a gas scale height. Taking $f = \sigma / \sigma_*$ interpolates between two extreme cases: when  $\sigma/\sigma_* \ll 1$, the scale height should approach the single-component value, i.e. $f=0$. When $\sigma\sim \sigma_*$, the two-component disk behaves (at least in terms of vertical density) like a single fluid with surface density $\Sigma + \Sigma_*$, i.e. $f=1$. Note that the stellar scale height, which does not directly affect the dynamics of the disk, is just taken to be the single component solution,
\begin{equation}
H_* = \frac{\sigma_*^2}{\pi G (\Sigma + \Sigma_*)},
\end{equation}
which is reasonable for the small values of $f_g$ found within the star-forming regions of the disk. In reality the vertical structure of a self-gravitating disk in a dark matter halo is not this simple. However, excluding the effects of dark matter introduces an error of only 13\%, even in the dark-matter dominated regions of the outer disk \citep{Narayan:2002cj}. Given the uncertainty in $\eta$, this approximation is adequate.} 

\textcolor{black}{Substituting for the scale height and $\sigma_{nt}^2=\sigma^2-\sigma_t^2$, we obtain a radiative loss rate of}
\begin{equation}
\mathcal{L} = \eta \Sigma \sigma^2 \textcolor{black}{\kappa Q_g^{-1}  \left(1 + \frac{\Sigma_*\sigma}{\Sigma\sigma_*}  \right)} \left(1 - \frac{\sigma_{t}^2}{\sigma^2}  \right)^{3/2}
\end{equation}
\textcolor{black}{In this form, the radiative loss rate is the gas kinetic energy per dynamical time multiplied by a factor to account for the effect of stars on the disk's thickness and a factor to zero out the radiative losses when there is no turbulence.}
As the gas velocity dispersion falls towards the constant thermal velocity dispersion, non-thermal motions die away, the gas no longer dissipates its energy via shocks, and $\mathcal{L}\rightarrow 0$. The gas temperature used to calculate $\sigma_t$ is a free parameter of the model, but fiducially we assume $T_g=7000 K$, appropriate for the warm neutral medium of the Milky Way. At high redshift when the gas is virtually all molecular, $T\ll 7000 K$, but in that regime $\sigma_t/\sigma \ll 1$ anyway, even if we use the higher-than-appropriate gas temperature. The choice of $\sigma_t$ therefore has virtually no effect on the high-redshift evolution of the disk.

\textcolor{black}{The governing equations for the gas (equations \ref{eq:dcoldt} and \ref{eq:dsigdt}) are derived under the assumption that $v_z=0$. We therefore implicitly neglect the gravitational potential energy of the disk associated with its vertical extent, and the associated $P\, dV$ work that the gas performs when it changes its scale height. Qualitatively, the effect of including these terms would be to provide the gas with another place to store energy which it gains when falling down the galaxy's potential well, aside from turbulent motion. Thus with these effects the gas velocity dispersion would be slightly lower, and hence so would the dissipation rate, the gas column density, and the star formation rate. The dissipation rate and star formation rate are each already controlled by a free parameter which is uncertain at the factor of two level, so we are content to neglect these additional repositories of energy. }
\subsection{Gravitational Instability}
\label{torque}
The stability against gravitational collapse of a self-gravitating disk is given by a Toomre Q-like parameter. Several such fragmentation conditions exist in the literature. We adopt \textcolor{black}{a modified version of} the condition determined by \citet{Rafikov:2001vp}, wherein the stability of a multi-component disk is considered with the stars treated realistically as a collisionless fluid.
\begin{equation}
Q(q)^{-1} = Q_g^{-1} \frac{2q}{1+q^2} + \sum_{i}\left[Q_{*,i}^{-1} \frac{2}{q \phi_i}\left( 1- e^{-q^2 \phi_i^2} I_0(q^2 \phi_i^2  \right)  \right]
\end{equation}
where $i$ indexes an arbitrary number of stellar populations, $\phi_i$ is the ratio of the ith stellar population's velocity dispersion to the gas velocity dispersion, $I_0(x)$ is a modified Bessel function of the first kind, and the Q parameter for each component is defined by
\begin{equation}
Q_j = \frac{\kappa \sigma_j}{\pi G \Sigma_j}.
\end{equation}
The epicyclic frequency is $\kappa = \sqrt{2(\beta+1)} \Omega $, and $q= k\sigma/\kappa$ is the dimensionless wavenumber, where k is the dimensional wavenumber of the perturbation. Values of q, or equivalently k, for which $Q(q) < 1$ are unstable \textcolor{black}{for an infinitely thin disk}, and the q which minimizes $Q(q)$ corresponds to the least stable mode. It follows that if $Q\textcolor{black}{_{Raf}}=\mbox{min}(Q(q)) < 1$, the disk is \textcolor{black}{formally} unstable, while if $Q\textcolor{black}{_{Raf}}>1$, the disk is stable.

Computing the value of $Q$ requires a minimization with respect to $q$. Since $Q$ and its partial derivatives must be calculated frequently (see equation \ref{eq:torque1} below), it is computationally expedient to use an approximate formula which does not require such a minimization. KB10 used $Q^{-1}\approx Q_{WS}^{-1} \equiv Q_g^{-1} + Q_*^{-1}$, as proposed by \citet{Wang:1994ar}, but this approximation becomes inaccurate when $\sigma_g/\sigma_* \lesssim 0.5$. \citet{Romeo:2011re} have proposed a more accurate approximation
\begin{mathletters}
\begin{eqnarray}
&Q_{RW}^{\textcolor{black}{-1}} = & \left\{ \begin{array}{cc}
	\textcolor{black}{\frac{W}{Q_* T_*} + \frac{1}{Q_g T_g}} & \mbox{ if \textcolor{black}{$Q_* T_*\ge Q_g T_g$ }}, \\
	\textcolor{black}{\frac{1}{Q_* T_*} + \frac{W}{Q_g T_g}} & \mbox{ if \textcolor{black}{$Q_* T_*\le Q_g T_g$ }};
		\end{array} \right. \\
&W = &  \frac{2\sigma_*\sigma}{\sigma_*^2 + \sigma^2}.
\end{eqnarray}
\end{mathletters}
\textcolor{black}{This formula includes corrections for the fact that the disk is not razor-thin, $T_*$ and $T_g$. A disk of finite thickness is more stable against gravitational collapse because its mass is spread out vertically, so larger values of the $T_j$ increase the value of $Q$ for a given set of column densities and velocity dispersions. \citet{Romeo:2011re} give approximations to these correction factors, $T_j\approx 0.8 + 0.7 \sigma_{z,j}/\sigma_{r,j}$. For simplicity we have assumed an isotropic velocity dispersion, so $T_*=T_g=1.5$.}  $Q_{RW}$ and its partial derivatives are straightforward to compute and accurate over a wide range of $\sigma_g/\sigma_*$ and $Q_g/Q_*$. \textcolor{black}{The stability parameter as determined by $Q_{Raf}$ should also be modified to include the effects of disk thickness, so our code can use either $Q\approx Q_{Raf}T$ or $Q \approx Q_{RW}$.}

Disks where gravitational instability dominates the dynamics are expected to be self-regulated near $Q=1$ \citep{Burkert:2010kr}. A disk with $Q\lesssim 1$ develops inhomogeneities in the gravitational field, which exert random forces on gas in the disk, driving turbulence. The ultimate source of this energy is the gravitational potential of the galaxy, so mass must move inwards. If the disk had $Q\textcolor{black}{\lesssim}1$, more mass would gather into inhomogeneities, thereby increasing the driving of turbulence, which stabilizes the disk, driving Q upwards. Meanwhile if $Q\textcolor{black}{\gtrsim}1$, mass transport \textcolor{black}{through the disk} slows even if the \textcolor{black}{cosmological} accretion rate does not, which tends to add mass and destabilize the disk. We therefore take as a hypothesis that $Q$ \textcolor{black}{ is a constant of order unity} at all points in the disk at all times. Thus we can set
\begin{eqnarray}
\label{eq:torque1}
\frac{dQ}{dt} &=&\frac{\partial Q}{\partial \Sigma} \frac{\partial\Sigma}{\partial t} + \frac{\partial Q}{\partial \sigma} \frac{\partial\sigma}{\partial t} \nonumber \\
& &+ \sum_i \left( \frac{\partial Q}{\partial \Sigma_{*,i}} \frac{\partial\Sigma_{*,i}}{\partial t} + \frac{\partial Q}{\partial \sigma_{*,i}} \frac{\partial\sigma_{*,i}}{\partial t} \right) = 0. 
\end{eqnarray}
The evolution of the gas state variables $\Sigma$ and $\sigma$, derived in the previous section, depends on the viscous torque and its radial derivatives, so we can recast equation \eqref{eq:torque1} in the form
\begin{equation}
\label{torqueeq2}
\frac{dQ}{dt} = f_2 \frac{\partial^2 \mathcal{T}}{\partial r^2} + f_1 \frac{\partial \mathcal{T}}{\partial r} + f_0 \mathcal{T} - F = 0,
\end{equation}
where the $f_i$ are coefficients which can be read off from the gas evolution equations, and F encompasses all terms which do not depend on the viscous torque, including all stellar processes, discussed in the following section, and the rate of radiative dissipation. In particular,
\begin{mathletters}
\begin{eqnarray}
f_2 & = & -\frac{\sigma}{6\pi r\Sigma v_\phi (\beta+1)} \frac{\partial Q}{\partial \sigma} - \frac{1}{2\pi(\beta+1) r v_\phi} \frac{\partial Q}{\partial \Sigma}, \\
f_1 & = & \frac{\beta^2\sigma + \sigma (r \frac{\partial\beta}{\partial r}+\beta) - 5(\beta+1)r\frac{\partial \sigma}{\partial r}}{6\pi(\beta+1)^2 r^2 v_\phi \Sigma} \frac{\partial Q}{\partial \sigma} \nonumber \\
& &+  \frac{\beta(\beta+1) + r (\partial\beta/\partial r)}{2 \pi (\beta + 1)^2 r^2 v_\phi}  \frac{\partial Q}{\partial \Sigma}, \\
f_0 & = & \frac{1}{6\pi r \Sigma}  (\beta-1)\frac{v_\phi}{r^2\sigma} \frac{\partial Q}{\partial \sigma}, \\
\label{forcing}
F & = & \textcolor{black}{\frac{\eta \pi}{3} G\Sigma\left(1 + \frac{\Sigma_*\sigma}{\Sigma\sigma_*}  \right)\left(1 - \frac{\sigma_t^2}{\sigma^2}\right)^{3/2}} \frac{\partial Q}{\partial \sigma} \nonumber \\ 
& &+ (f_R+\mu)\dot{\Sigma}_*^{SF} \frac{\partial Q}{\partial \Sigma} \nonumber \\
& &- \sum_i\left( \dot{\Sigma}_{*,i}\frac{\partial Q}{\partial \Sigma_{*,i}} + \dot{\sigma}_{*,i} \frac{\partial Q}{\partial \sigma_{*,i}}\right).
\end{eqnarray}
\end{mathletters}
Usually F will be dominated by the first term, the radiative dissipation of energy, which tends to destabilize the disk by ``cooling'' the gas, making $F>0$. In this case, one can interpret equation \eqref{torqueeq2} as requiring the torques to move gas such that it stabilizes the disk to counter the effects of this cooling. 

Equation \eqref{torqueeq2} is a second order ODE requiring two boundary conditions. At the outer edge of the disk, we specify the accretion rate of gas onto the disk, $\dot{M}_{ext}$ according to a pre-calculated accretion history, typically a fit to average accretion histories from cosmological simulations \citep{Bouche:2010xz}. The torque is related to $\dot{M}$ through equation \eqref{mdot}, so by rearranging that equation, evaluating quantities at the outer radius, and requiring a particular $\dot{M}_{ext}$, we obtain the outer boundary condition
\begin{equation}
\left(\frac{\partial \mathcal{T}}{\partial r}\right)_{r=R} = -\dot{M}_{ext} v_\phi(R) (1+\beta(R)).
\end{equation}
Here $R$ is a fixed outer radius of the disk. This condition implicitly assumes that all gas is accreted at the outer edge of the disk, which is not an unreasonable approximation as long as gas accretes mostly through cold streams

At the inner boundary, we require that the disk and bulge exert no torques on each other,
\begin{equation}
\label{eq:IBC}
\left(\mathcal{T}\right)_{r=r_0}=0
\end{equation}
The inner edge of the computational domain is $r_0$, chosen for numerical reasons to be non-zero. Note that this boundary condition is somewhat different than the one used in KB10, namely $(\mathcal{T})_{r=r_0} = -\dot{M}_{ext} v_\phi(R)(1+\beta(R)) r_0$ for a flat rotation curve. This will approach the physically motivated value of equation \eqref{eq:IBC} in the limit that $r_0\rightarrow 0$, and was chosen to satisfy a regularity condition at the inner boundary. However, since our goal here is not to obtain an analytic solution, there is no need to impose such a condition.  In practice we have experimented with both choices in our numerical calculations, and we find that the choice of inner boundary condition has negligible effects at radii $r \gg r_0$, which is the great majority of the disk.

\section{Stellar Evolution Equations}
In addition to the gas, we would like to know how stellar populations in the disk evolve with time. The stars will provide most of the observable consequences of the model, in addition to determining, along with the gas, whether the disk is gravitationally unstable. Among the questions we are interested in investigating is the cause of the age-velocity dispersion correlation, namely that older stars have higher velocity dispersions. Therefore it is useful to not only keep track of the stars as a single population with a single column density $\Sigma_*$ and velocity dispersion $\sigma_*$ (each a function of radius and time), but also to bin the stars by age, so that $\Sigma_{*,i}$ and $\sigma_{*,i}$ describe the ith age bin.

The overall stellar population, along with each sub-population, will be directly affected by two processes - star formation and stellar migration. The two effects may be added together, recalling that of the gas which forms stars, only a fraction $f_R$ will remain in stars after stellar evolution has taken its course, 
\begin{equation}
\dot{\Sigma}_{*,i} = f_R \dot{\Sigma}_{*,i}^{SF} + \dot{\Sigma}_{*,i}^{Mig}.
\end{equation}
Evolution equations for each process will be derived separately below.

\subsection{Star Formation}
The rate of star formation will depend on the properties of the gas from which stars form. In particular, in a sufficiently large region of the disk, the star formation rate will be proportional to the molecular gas mass divided by the free fall time, defined to be $\sqrt{\textcolor{black}{3\pi/(32} G\rho)}$. In deriving the gas evolution equations, we assumed that formally the density was given by $\Sigma \delta(z)$, but this is of course an approximation. The disk will have a finite thickness of order the scale height (defined by equation \ref{eq:scaleheight}), so we take the density to be $\rho = \Sigma /H$. Thus we can write the star formation rate density
\begin{equation}
\label{sfr}
\dot{\Sigma}_*^{SF} = \epsilon_{\mbox{ff}} f_{H_2}\Sigma \sqrt{\textcolor{black}{32}G\rho\textcolor{black}{/(3\pi)}} =\textcolor{black}{\epsilon_{\mbox{ff}} f_{H_2} \Sigma\kappa \sqrt{32/3}\left(1+\frac{\Sigma_*}{\Sigma}\frac{\sigma}{\sigma_*}\right)^{1/2}}
\end{equation}
For molecular gas, the efficiency of star formation per free-fall time is $\epsilon_{\mbox{ff}} \textcolor{black}{\sim} 0.01$ \citep{Krumholz:2005vz, Krumholz:2007qw, Krumholz:2011dm}, though this may be significantly higher or lower given observational uncertainties. Following progress made by \citet{Krumholz:2008mj,Krumholz:2009mj}, \citet{2010ApJ...709..308M} have analytically approximated the molecular fraction of the gas, $f_{H_2}$ as a function of metallicity and surface density. We adopt this prescription with a slight alteration:
\begin{mathletters}
\begin{eqnarray}
f_{H_2} &=&\left\{ \begin{array}{cc}
	1 - \left(\frac{3}{4} \right)\frac{s}{1 + 0.25 s} &\mbox{ if $s < 388/203$} \\
	0.03 &\mbox{ otherwise} \\
		\end{array} \right. \\
s & = & \frac{\ln(1 + 0.6\chi + 0.01\chi^2)}{0.6 \tau_c} \\
\chi & = & 3.1\ \frac{1 + 3.1 (Z/Z_\odot)^{0.365}}{4.1} \\
\tau_c & = & 320\ c\ (Z/Z_\odot) (\Sigma / 1\mbox{ g cm}^{-2}),
\end{eqnarray}
\end{mathletters}
where Z is the gas metallicity. We take the solar metallicity to be $Z_\odot = 0.02$, and $c$ encapsulates the effects of clumping in the gas when averaging over large regions. Since the model presented in this paper takes averages over large areas of the disk, we take $c\sim 5$, as determined in \citet{Krumholz:2009lt}. The modification from \citet{2010ApJ...709..308M} is that we impose a lower limit on $f_{H_2}$ of $3\%$, motivated by the observation that even extremely low total gas surface density regions form stars at a rate consistent with a constant H$_2$ depletion time \citep{2011ApJ...730L..13B}. 

Equation \eqref{sfr} is used to update the stellar column density, and it also enters into the gas column density equation (equation \ref{eq:dcoldt}) through the conservation of mass. At any particular time in a simulation, all but one of the $\dot{\Sigma}_{*,i}^{SF} = 0$. Formally we can write this as
\begin{equation}
\dot{\Sigma}_{*,i}^{SF} = \dot{\Sigma}_*^{SF}\ \Theta(A(t) - A_{young,i})\ \Theta(A_{old,i} - A(t))
\end{equation}
where $\Theta(x)$ is a step function, one for $x>0$ and zero for $x<0$, $A(t)$ is the age that a star will be at redshift zero if it forms at time t after the beginning of the simulation, and $A_{young,i}$ and $A_{old,i}$ are the boundaries of the ith age bin.

To update the stellar velocity dispersion of a stellar population, we require that the new kinetic energy of the population be equal to the old kinetic energy plus the energy of the newly formed stars,
\begin{equation}
\label{energycons}
(\Sigma_{*,i} \sigma_{*,i}^2)_{new} = (\Sigma_{*,i}\sigma_{*,i}^2)_{old} + f_R (d\Sigma_{*,i}^{SF})\sigma^2
\end{equation}
where we have assumed that the newly formed stars have the same velocity dispersion as the gas from which they form. Setting $\Sigma_{*,new}= \Sigma_{*,old} + f_R(d\Sigma_*^{SF})$, we can rearrange, solve for $\sigma_{*,new}$, and expand to first order in the small quantity $d\Sigma_*^{SF}/\Sigma_{*,old}$
\begin{eqnarray}
\sigma_{*,i,new} & = & \sqrt{ \frac{(\Sigma_{*,i}\sigma_{*,i}^2)_{old} + f_R(d\Sigma_{*,i}^{SF})\sigma^2}{\Sigma_{*,i,old} + f_R (d\Sigma_{*,i}^{SF}) } } \nonumber \\
& \approx & \sigma_{*,i,old} + \frac{f_R (d\Sigma_{*,i}^{SF})}{2 \Sigma_{*,i,old}\sigma_{*,i,old}} (\sigma^2 - \sigma_{*,i,old}^2)
\end{eqnarray}
Thus in the limit that the time step and therefore the density of new stars produced is small, we may use the definition of a derivative to write
\begin{equation}
\label{eq:dsigstdt}
\left(\frac{\partial \sigma_{*,i}}{\partial t}\right)^{SF} \approx f_R\ \frac{1}{2\Sigma_{*,i}\sigma_{*,i}}(\sigma^2 - \sigma_{*,i}^2) \dot{\Sigma}_{*,i}^{SF}\ \ \ \ \mbox{for   } \Sigma_{*,i}>0
\end{equation}
We only need this derivative for its contribution to the torque equation (equation \ref{eq:torque1}), in which it will always be multiplied by the term $\partial Q/\partial\sigma_{*,i}$. To actually update the quantity $\sigma_{*,i}$, we use the exact relation of equation \eqref{energycons}, which holds even if $\Sigma_{*,i}=0$. Note that when $\Sigma_{*,i}=0$, this new population of stars will have no effect on the torque equation, since $\partial Q/\partial\sigma_{*,i} = 0$, i.e. non-existent stars do not affect the stability of the disk. Thus equation \eqref{eq:dsigstdt} need only be employed when $\Sigma_{*,i}>0$.

\subsection{Radial Migration}
\label{sec:stmig}
In addition to star formation, stars are subject to radial migration. In particular, when $Q_* \lesssim 2$, transient spiral arms form which attempt to stabilize the disk \citep{Sellwood:1984mb, Carlberg:1985ri, Sellwood:2002kk}. N-body simulations \citep{Sellwood:1984mb} suggest that this heating is such that
\begin{equation}
\label{eq:heating}
{\frac{\partial Q_*}{\partial t}}^{Mig} = \mbox{max} \left( \frac{Q_{lim} - Q_*}{T_{mig}  (2\pi \Omega^{-1})}, 0 \right)
\end{equation}
where $T_{mig}$ is the time scale in local orbital times over which this heating occurs, typically a few orbits, and $Q_{lim}$ is the value of $Q_*$ above which the stars are stable to spiral perturbations. \textcolor{black}{Equation \eqref{eq:heating} assumes that this mechanism acts independently of the torques which act on the gas as a result of the axisymmetric instability described in section \ref{torque}. In $z\sim 2$ galaxies with morphologies dominated by clumps containing both gas and stars, one might expect the axisymmetric instability to affect both components equally, as assumed in the models of \citet{Cacciato:2011ab}. However, it remains an open question whether these clumps are disrupted on a dynamical timescale by a stellar feedback process, just like giant molecular clouds, their low-redshift analogues \citep{Krumholz:2010ad,Genel:2012ng}.  Even if clumps are long-lived, they contain a relatively small part of the total stellar population \citep{Murray:2010tg}, and thus their impact on stellar migration might be small. Moreover, in most realistic situations, the scale height of stars will be significantly greater than that of the gas, so an instability dominated by the gas will have little traction on the stars. As long as $\sigma_*/\sigma$ is appreciably greater than unity, which it is in our fiducial model (section \ref{sec:fid}), we expect this treatment to be reasonable.}

The time derivative \textcolor{black}{of $Q_*$} may be re-expressed \textcolor{black}{in terms of the time derivatives of $\Sigma_*$ and $\sigma_*$} using the definition of $Q_*$,
\begin{eqnarray}
\label{product}
{\frac{\partial Q_*}{\partial t}}^{Mig} & = & \frac{\kappa }{\pi G} \left(\frac{1}{\Sigma_*} {\frac{\partial \sigma_*}{\partial t}}^{Mig} -\frac{\sigma_*}{\Sigma_*^2}{\frac{\partial \Sigma_*}{\partial t}}^{Mig} \right) \nonumber \\
& = & Q_* \left( \frac{1}{\sigma_*}{\frac{\partial \sigma_*}{\partial t}}^{Mig} - \frac{1}{\Sigma_*} {\frac{\partial \Sigma_*}{\partial t}}^{Mig} \right)
\end{eqnarray}
\textcolor{black}{The partial} time derivatives \textcolor{black}{on the right hand side} will depend on the mean velocity of stars in the radial direction, $v_{*,r}$, and so the \textcolor{black}{forcing} imposed by equation \eqref{eq:heating} will yield an ordinary differential equation for $v_{*,r}$\textcolor{black}{. The value of $v_{*,r}$ is then used to evolve $\Sigma_*$ and $\sigma_*$.}

\textcolor{black}{This approach assumes a single bulk velocity of stars in the radial direction at each radius, $v_{*,r}(r)$. It has been well-demonstrated \citep[e.g.][]{Bird:2012nb} that the \citet{Sellwood:2002kk} mechanism scatters stars in both directions, i.e. a star born at some galactocentric radius may end up with a guiding center radius multiple kpc away. There are additional scattering mechanisms, such as two-body scattering and the resonant overlap between spirals and the bar \citep{Minchev:2010bf, Brunetti:2011cp}, which will also redistribute stellar angular momenta. Modeling this redistribution is critical in explaining the detailed properties of Milky Way stellar populations. However, there are no straightforward prescriptions to model all of these effects. We therefore ignore for now the effects of radial mixing and merely require conservation of mass and energy, and that the stars will stabilize themselves if they are subject to spiral instabilities.}


\textcolor{black}{The evolution of $\Sigma_*$ and $\sigma_*$ as a function of $v_{*,r}$ is determined by the continuity equations for mass and energy of the ith stellar population} 
\begin{equation}
\label{eq:stMassCont}
{\frac{\partial \Sigma_{*,i}}{\partial t}}^{Mig} + \frac{1}{r}\frac{\partial}{\partial r}(r\Sigma_{*,i} v_{*,r}) = 0
\end{equation}
\begin{eqnarray}
\frac{\partial }{\partial t}\left[\Sigma_{*,i}\left(v_\phi^2+3\sigma_{*,i}^2+2\psi\right)\right]^{Mig} + & &\nonumber \\
\frac{1}{r}\frac{\partial}{\partial r}\left[r\Sigma_{*,i} v_{*,r}\left(v_\phi^2+3\sigma_{*,i}^2+2\psi\right)\right] &=& 0
\end{eqnarray}

Expanding the energy equation using the product rule and employing the mass equation to cancel terms leaves
\begin{eqnarray}
\Sigma_{*,i}\frac{\partial }{\partial t}\left[\left(v_\phi^2+3\sigma_{*,i}^2+2\psi\right)\right] + & &\nonumber \\
\Sigma_{*,i} v_{*,r}\frac{\partial}{\partial r}\left[\left(v_\phi^2+3\sigma_{*,i}^2+2\psi\right)\right]&=& 0
\end{eqnarray}
The time derivatives of $v_\phi$ and $\psi$ are zero by assumption, so expanding the surviving derivatives, setting $\partial \psi/\partial r = v_\phi^2/r$ and $\partial v_\phi / \partial r = \beta v_\phi / r$, and rearranging yields
\begin{equation}
{\frac{\partial \sigma_{*,i}}{\partial t}}^{Mig} = - v_{*,r} \left(\frac{(1+\beta)v_\phi^2}{3r\sigma_{*,i}} + \frac{\partial \sigma_{*,i}}{\partial r} \right)
\end{equation}
The corresponding equation for stellar column density follows immediately from the continuity equation:
\begin{equation}
{\frac{\partial \Sigma_{*,i}}{\partial t}}^{Mig} =-\Sigma_{*,i} \frac{\partial v_{*,r}}{\partial r} - v_{*,r}\frac{\partial \Sigma_{*,i}}{\partial r} - \Sigma_{*,i} v_{*,r}/r
\end{equation}
Substituting the transport equations into equation \eqref{product} and imposing equation \eqref{eq:heating} yields
\begin{eqnarray}
\label{stmig}
2\pi r \frac{v_{*,r}}{v_\phi}  \left( -\frac{v_\phi^2}{\sigma_*^2}\frac{(1+\beta)}{3r} - \frac{1}{\sigma_*}\frac{\partial \sigma_*}{\partial r}    + \frac{1}{\Sigma_*}\frac{\partial \Sigma_*}{\partial r} +  1/r  \right) \nonumber \\
+  \frac{2\pi r}{v_\phi}\frac{\partial v_{*,r}}{\partial r} =\frac{\mbox{max}(Q_{lim}-Q_*,0)}{T_{Mig} Q_*} 
\end{eqnarray}
This is a first order ordinary differential equation (since at any particular time we treat all variables as functions of radius only), requiring a single boundary condition which we take to be $v_{*,r} (r=R) = 0$, which means that no stars are allowed to migrate between the outer edge of the disk and the IGM. \textcolor{black}{This boundary condition guarantees that the bulk velocity of stars in the radial direction will be inwards at all radii, which means this method does not conserve angular momentum; to compensate for a large mass of stars moving inwards, a small mass of stars would need to move outwards. The error we make in conservation of total angular momentum is about 2\% in the fiducial case.} 


\section{Metallicity Evolution}
\label{sec:metallicity}
\subsection{\textcolor{black}{Advection of Metals in Gas}}
To describe the evolution of the metal content, we begin by defining $\Sigma_Z$, the surface density of metals, so locally the metallicity of the gas is $Z=\Sigma_Z/\Sigma$. The continuity equation for $\Sigma_Z$ is
\begin{equation}
\label{metalcontinuity}
\frac{\partial}{\partial t} \Sigma_Z = \frac{1}{2\pi r}\frac{\partial}{\partial r}\dot{M}_Z  -\dot{\Sigma}_{Z}^{SF} + S_Z
\end{equation}
where $\dot{\Sigma}_{Z}^{SF} = \dot{\Sigma}_*^{SF} Z$ is the rate at which metals are incorporated into newly formed stars, and $S_Z$ is a source term for metals injected into the gas by supernovae and AGB stars. Note that, in writing this equation, we neglect transport of metals through the disk by either turbulent diffusion or galactic fountains. The inward flux of metallic mass  is
\begin{equation}
\dot{M}_Z = \dot{M} Z = -\frac{Z}{v_\phi (1+\beta)} \frac{\partial \mathcal{T}}{\partial r},
\end{equation}
which follows from equation \ref{mdot}. The left hand side of equation \ref{metalcontinuity} can be reexpressed in terms of Z by noting $\partial \Sigma_Z/\partial t = Z \partial \Sigma/\partial t + \Sigma \partial Z / \partial t$. Equation \ref{metalcontinuity} then becomes
\begin{eqnarray}
Z\frac{\partial \Sigma}{\partial t} +\Sigma \frac{\partial Z}{\partial t} &=& \nonumber \\
\frac{Z}{2 \pi  r (1+\beta)^2 v_\phi}  &\Bigg(& (1+\beta) \frac{\partial \mathcal{T}}{\partial r} \beta /r - (1+\beta) \frac{\partial \mathcal{T}}{\partial r} \frac{\partial \ln Z}{\partial r} \nonumber \\
+ \frac{d\beta}{dr} &\frac{\partial \mathcal{T}}{\partial r}&-(1+\beta) \frac{\partial^2 \mathcal{T}}{\partial r^2}\Bigg) - \dot{\Sigma}_Z^{SF} + S_Z
\end{eqnarray}
Comparing this with the previously derived gas surface density evolution equation, we can cancel most of the terms on the right hand side with $Z \partial \Sigma/\partial t$, leaving only
\begin{equation}
\label{eq:zadvec}
\frac{\partial Z}{\partial t} = -\frac{1}{(\beta+1) r\Sigma v_\phi} \frac{\partial \ln Z}{\partial r}\frac{\partial \mathcal{T}}{\partial r} + \frac{S_Z}{\Sigma}.
\end{equation}
Inflowing gas has some metallicity $Z_{IGM}$, which we fix at $Z_{IGM}=0.1Z_\odot$ for the entire simulation. Simulations \citep{Shen:2011er} and observations \citep{Adelberger:2005zl} suggest that the circum-galactic medium is enriched to this degree as early as $z=3$. 

\subsection{\textcolor{black}{Metal Production}}
For simplicity, we adopt the instantaneous recycling approximation, proposed by \citet{1980FCPh....5..287T}, to specify $S_Z$, the production rate of metals. First we recognize that metals are produced in supernovae and AGB stars. To a first approximation, we can assume that the lifetimes of stars that dominate metal production are much smaller than the timescales with which we are concerned in this paper, so metals enter the ISM at a rate proportional to the star formation rate. Not all gas which forms stars is returned to the ISM, since low-mass stars do not leave the main sequence in a Hubble time and even high-mass stars form remnants. Defining the remnant fraction $f_R$ as the fraction of gas forming stars which will end up not being returned to the ISM, the surface density of recycled gas appearing in the ISM is $(1-f_R)\dot{\Sigma}_*^{SF}$. Supernovae and normal stellar evolution will enrich a small fraction of this gas, namely $y_M$, the yield.  The surface density of metal production is therefore
\begin{equation}
S_Z = y_M \textcolor{black}{\zeta}(1-f_R)\dot{\Sigma}_*^{SF}.
\end{equation}
Assuming a \citet{Chabrier:2005wm} initial mass function and a coarse approximation for the ultimate fates of stars as a function of mass, \citet{2011arXiv1106.0301K} compute $f_R=0.46$. Assuming in addition a production function, the fraction of a star's initial mass converted to a given element, from \citet{1992A&A...264..105M}, they compute a yield of $y_M=0.054$ for Solar metallicity stars. The effective yield may be somewhat smaller than this, since galactic winds driven by supernovae tend to eject material which is richer than average in metals. \textcolor{black}{The factor of $\zeta \lesssim 1$ represents the ratio of metallicity in the ISM to metallicity of ejected material. We adopt $\zeta=1$, corresponding to an assumption that the ejecta are well-mixed with the ISM. This value will in principle depend on the mass of the galaxy considered \citep{MacLow:1999af}. However, owing to the high resolution required, to date no simulation has reliably calculated the degree to which metal-rich gas is preferentially ejected. Changes in the exact value of $\zeta$ roughly translate into the normalization of the metallicity distribution in the gas, so our fiducial value of $\zeta$ was chosen to give reasonable values for this normalization.}

\subsection{\textcolor{black}{Diffusion of Metals}}
\textcolor{black}{The metallicity gradients produced when accounting only for metal production by stars and advection by inflowing gas are far steeper than the observed gradient in the Milky Way. Metals are formed in proportion to the star formation rate, which tends to be high towards the center of the simulated galaxies. Meanwhile the inflow of gas throughout the disk concentrates the metals even further. To explain the relatively small observed metallicity gradients, one must allow metals formed at small galactic radii to reach large radii. This may occur either in the plane of the disk (diffusion) or out of the plane (galactic fountains). By assuming a fixed value of $Z_{IGM} = 0.1 Z_\odot$, we have already implicitly assumed some sort of transport of metals from the galaxy into its surrounding medium. However, rather than modeling this transport in any detail, let us consider only the diffusion of metals through the disk.}

\textcolor{black}{In general, a diffusion equation will have the form
\begin{equation}
\frac{\partial}{\partial t} M_Z = D \frac{\partial^2}{\partial r^2} M_Z
\end{equation}
where $D$ is the diffusion coefficient and $M_Z = 2\pi r \Delta r \Sigma Z$ is the gas-phase metal mass in a given cell. At an order of magnitude level $D$ may be estimated by taking the typical velocity of gas in the disk, $\sigma$, and multiplying by the typical length scale of perturbations, namely the 2d Jeans scale, $\sigma^2/G\Sigma$. For simplicity we simply adopt $D = k_Z v_\phi(R) R$ where $k_Z$ and $D$ will be constant at every time and location in the disk. Numerically we take $k_Z = 10^{-3}$ which is of the correct order of magnitude and yields a metallicity gradient of order $0.1$ dex/kpc (see figure \ref{fig:rtsc})., which is comparable to observed values in isolated spiral galaxies \citep[e.g.][]{Zaritsky:1994kh}}

\subsection{\textcolor{black}{Metals Locked in Stars}}
The metallicity of a given stellar population can be updated \textcolor{black}{when new stars are added to it by again assuming instantaneous missing.}The new metallicity is just an average of the old metallicity and the metallicity of the gas, weighted by the surface density of \textcolor{black}{the extant stellar population and the newly formed population respectively.}
\begin{equation}
\label{eq:stzform}
Z_{*,i,new} = \frac{Z_{*,i,old} \Sigma_{*,i} + f_R Z (d\Sigma_{*,i}^{SF}) }{\Sigma_{*,i} + f_R (d\Sigma_{*,i}^{SF})}
\end{equation}
\textcolor{black}{Here, as in equation \eqref{energycons}, $d\Sigma_{*,i}^{SF} = \dot{\Sigma}_{*,i}^{SF} dt$, is the surface density of stars formed in a given time step in a given stellar age bin $i$.}

\textcolor{black}{Besides the formation of new stars, a given stellar population is subject to migration through the disk, as discussed in section \ref{sec:stmig}} Since the stars migrate through the disk with a mean velocity set by equation \eqref{stmig}, the metallicity profile of a given population of stars evolves under a continuity equation for the metal mass,
\begin{equation}
\frac{\partial}{\partial t} \left( \Sigma_{*,i} Z_{*,i} \right)^{Mig} + \frac{1}{r}\frac{\partial}{\partial r} \left(r\Sigma_{*,i}Z_{*,i} v_{*,r}  \right) = 0
\end{equation}
Subtracting the continuity equation for total stellar mass \textcolor{black}{(equation \ref{eq:stMassCont})}, we obtain
\begin{equation}
\label{eq:stzmig}
\frac{\partial Z_{*,i}}{\partial t}^{Mig} = -v_{*,r} \frac{\partial Z_{*,i}}{\partial r},
\end{equation}
for the evolution of stellar metallicity. \textcolor{black}{Equations \eqref{eq:stzform} and \eqref{eq:stzmig} fully describe the evolution of the metallicity of the ith stellar population.} Note that \textcolor{black}{these equations neglect} radial diffusion of stars, only taking into account the mean velocity $v_{*,r}$. Radial mixing \citep{Sellwood:2002kk,Roskar:2011rm} is required to explain the spread of metallicities in stars at a fixed age and radius, and undoubtedly leads to a shallower stellar metallicity gradient than what we obtain. 

\section{Numerical Method}

\subsection{Computational Domain}
In deriving the gas evolution equations, we assumed the disk to be thin and axisymmetric. Thus the disk is described by variables which change only in radius and time. We therefore define a mesh of radial positions $r_i$ with a fixed number of points, $n_x$, logarithmically spaced between the outer edge of the disk at a fixed radius R and a fixed inner cutoff $r_{min}$, usually chosen to be $r_{min}= 0.01 R$. Explicitly,
\begin{equation}
r_i = R  \left(\frac{r_{min}}{R}\right)^{1-(i-1)/(n_x-1)}
\end{equation}
The highest spatial resolution is therefore given to the region with the shortest dynamical times.

Time, tracked in units of the orbital period at radius R, begins at zero when the simulations are started, typically at $z=2$, and reach a few tens of orbits at $z=0$, depending on the assumed radius and circular velocity. The size of the time steps are calculated by first determining all timescales defined by dividing each state variable at each position by its time derivative, picking out the minimum timescale, and multiplying it by a small number TOL, usually taken to be $10^{-4}$. Larger values of TOL lead to numerical instabilities near the inner boundary, which is especially susceptible to such issues because the local dynamical timescale in the disk is $\Omega^{-1} \propto r$ for a flat rotation curve.
\begin{eqnarray}
\Delta t &=& \mbox{TOL} \times \mbox{min}_{i} \ \Big[ \frac{\Sigma}{\partial \Sigma / \partial t} (r_i), \frac{\sigma}{\partial \sigma / \partial t} (r_i), \nonumber \\
& & \frac{\Sigma_*}{\partial \Sigma_* / \partial t} (r_i) , \frac{\sigma_*}{\partial \sigma_* / \partial t} (r_i), \frac{0.01}{\mbox{TOL}} \Big]
\end{eqnarray}
A maximum time step of $0.01$ outer orbits is imposed to prevent systems extremely close to equilibrium from advancing too quickly.
 
\subsection{PDEs}
At each time step, the code solves the equations in non-dimensionalized form (see appendix) in the following order. First, we solve equation \eqref{stmig} to determine $v_{*,r}$ at all radii. The equation is of the form $\mathcal{H} = h_0 v_{*,r} + \partial v_{*,r}/\partial r$ with
\begin{mathletters}
\begin{eqnarray}
\mathcal{H} & = & \frac{\mbox{max}(Q_{lim}-Q_*,0) v_\phi}{2\pi r T_{Mig} Q_*}\\
h_0 & = & -\frac{v_\phi^2}{\sigma_*^2} \frac{(1+\beta)}{3 r} - \frac{1}{\sigma_*}\frac{\partial \sigma_*}{\partial r} + \frac{1}{\Sigma_*}\frac{\partial \Sigma_*}{\partial r} + \frac{1}{r},
\end{eqnarray}
\end{mathletters}
The boundary condition specifies $v_{*,r}$ at the outer edge of the disk. Thus rewriting the radial derivative as a finite difference and employing a backwards Euler step, we can write an explicit update equation,
\begin{equation}
v_{*,r}(r_{i-1}) \approx \frac{v_{*,r}(r_i) - (r_i - r_{i-1})\mathcal{H}(r_{i-1})}{1 - (r_i-r_{i-1}) h_0(r_{i-1})} ,
\end{equation}
which we solve iteratively by starting with the specified value of $v_{*,r}(r_{nx}) = 0$ and moving inwards.

Using the value of $v_{*,r}$ along with the current values of the state variables, we calculate the coefficients of the torque equation (equation \ref{torqueeq2}). To solve the resultant linear PDE, we employ a similar finite difference method, which approximates
\begin{eqnarray}
\frac{\partial\mathcal{T}_i}{\partial r} &\approx& \frac{\mathcal{T}_{i+1}-\mathcal{T}_{i-1}}{r_{i+1}-r_{i-1}} \\
\frac{\partial^2\mathcal{T}_i}{\partial r^2} &\approx& \frac{1}{r_{i+1/2}-r_{i-1/2}}\left(\frac{\mathcal{T}_{i+1}-\mathcal{T}_{i}}{r_{i+1}-r_{i}} - \frac{\mathcal{T}_{i}-\mathcal{T}_{i-1}}{r_{i}-r_{i-1}}\right)
\end{eqnarray}
Since we are using a logarithmically spaced grid, $r_{i+1/2} = \sqrt{r_ir_{i+1}}$. By plugging these approximations into the torque equation, the problem reduces to the inversion of a tridiagonal matrix.

The forcing term in the torque equation, \eqref{forcing} generally acts to destabilize the disk, since its largest term comes from radiative cooling of the gas and cooler gas is more prone to gravitational collapse. The torque equation requires that the gravitational torques exactly counteract this effect to maintain $dQ/dt=0$. However, in the event that the forcing term in the torque equation becomes negative as a result of stellar migration and a reduced rate of cosmological infall leading to $\mathcal{L} \rightarrow 0$, we set it to zero so that the gas is not forced to destabilize the disk. This in turn allows positive values of $dQ/dt$. We do not allow the forcing term to return to the value given by \eqref{forcing} until that value is again positive and Q has been allowed to rise and then fall back down to $Q=\textcolor{black}{Q_f}$. This allows the simulation to follow disks which stabilize at least temporarily, for example because of a lull in the cosmological accretion rate, and then return to a marginally unstable state. For the smoothed average cosmological accretion history used in our fiducial run, parts of the disk which stabilize remain that way because the accretion rate is monotonically decreasing.

With $\mathcal{T}$, $\partial \mathcal{T}/\partial r$, $\partial^2\mathcal{T}/\partial r^2$, and $v_{*,r}$, we can now evaluate the derivatives of the state variables. Where radial derivatives of the state variables or other quantities appear in the evolution equations or the coefficients of the above differential equations, a minmod slope limiter is used to evaluate them. In particular, if $L = (A(r_i) - A(r_{i-1})) / (r_i - r_{i-1})$ and $R = (A(r_{i+1}) - A(r_i))/(r_{i+1}-r_i)$
\begin{equation}
\frac{\partial A}{\partial r} (r_i) = \left\{ \begin{array}{cc}
	L &\mbox{ if $| L | < | R |$ and $LR>0$} \\
	R &\mbox{ if $| L | > | R |$ and $LR>0$} \\
	0 &\mbox {otherwise}
		\end{array} \right.
\end{equation}
where A is a stand-in for any quantity. This strongly suppresses noise on the scale of the mesh separation by zeroing out rapid variations in the derivatives. 


With the time derivatives calculated at each point, we simply take a forward Euler step to update the state variables, namely $\Sigma$, $\sigma$, $Z$, $\Sigma_*$, $\sigma_*$, and for each age-binned stellar population, $\Sigma_{*,i}$, $\sigma_{*,i}$, and $Z_{*,i}$. Typical runs have time steps limited by the rate of change of the gas state variables near the inner boundary of the disk where the dynamical timescale is shortest. On a single processor, runs take about one day to complete if we numerically evaluate $Q(q)$ and its derivatives using the full \citet{Rafikov:2001vp} formalism. We can shorten this by an order of magnitude by using the approximation to $Q$ suggested by \citet{Romeo:2011re}. This approximation is much more efficient because $Q_{RW}$ and its partial derivatives may be calculated as functions of the state variables alone, without the need to minimize over a wavenumber or compute the partial derivatives $\partial Q/\partial \{\Sigma,\sigma,\Sigma_{*,i},\sigma_{*,i} \}$ numerically as required by the full Rafikov Q.

\subsection{Initial Conditions}
\label{sec:ic}
By assuming a flat rotation curve, fixed gas fraction, equal stellar and gas velocity dispersions, a simple analytic approximation to Q, and ignoring stellar processes (formation and migration), KB10 were able to compute an equilibrium solution to the evolution equations. In particular, 
\begin{eqnarray}
\label{eq:equilibrium_sig}
\sigma &=& \frac{1}{\sqrt{2}} \left( \frac{G\dot{M}_{ext,0}}{\eta f_g} \right) ^{1/3} \\
\label{eq:equilibrium_col}
\Sigma &=& \frac{v_\phi}{\pi G r} \left( \frac{f_g^2 G\dot{M}_{ext,0}}{\eta} \right) ^{1/3}
\end{eqnarray}
Here $\dot{M}_{ext,0}$ is the accretion rate of gas onto the outer edge of the disk at the start of the simulation, and $f_g$ is the gas fraction, assumed to be constant in radius. By assumption, $\sigma_* = \sigma$ and $\Sigma_* = \Sigma(1-f_g)/f_g$. 

If we relax the assumption\textcolor{black}{s} that the velocity dispersions of both components are identical and \textcolor{black}{$Q=1$, add a factor to correct for finite disk thickness, but retain the approximate form of $Q$ for an infinitely thin disk}, $Q^{-1}\approx Q_g^{-1}+Q_*^{-1}$, we obtain a modified version of the equilibrium column density,
\begin{equation}
\label{eq:initialize}
\Sigma = \textcolor{black}{\frac{T}{Q_f}} \frac{v_\phi}{\pi G r}\frac{\phi_0 f_g}{f_g(\phi_0 - 1)+ 1 }\left( \frac{G \dot{M}_{ext,0}}{\eta f_g} \right)^{1/3}
\end{equation}
where $\phi_0 = \sigma_*/\sigma$ is a free parameter \textcolor{black}{, $T\approx 1.5$ is the thickness correction, and $Q_f$ is the fixed value to which $Q$ will be set everywhere in the disk}. To initialize the simulations, we use equations \eqref{eq:equilibrium_sig} and \eqref{eq:initialize}. We then adjust $\sigma_*=\phi_0\sigma$ keeping $\phi_0$ fixed until $Q=\textcolor{black}{Q_f}$ exactly at each cell of the grid. Finally, we run the simulation with stellar processes turned off, i.e. $\epsilon_{ff}=Q_{lim}=0$, and with $\dot{M}_{ext}$ fixed to its initial value, $\dot{M}_{ext,0}$, to allow the gas to adjust to an equilibrium configuration. The greatest effect of this adjustment occurs at the inner edge of the disk, since these relations were derived using a different inner boundary condition and under a more stringent set of assumptions. Once the state variables are changing sufficiently slowly, we have found our initial conditions and therefore return $\epsilon_{ff}$, $Q_{lim}$, and $\dot{M}_{ext}(t)$ to their user-specified values.

\section{Fiducial Model}
\label{sec:fid}
While our code is quite general, here we describe a simple model run using it in order to demonstrate its capabilities. In future work we will explore a much wider part of parameter space, using more realistic cosmological accretion histories.
\subsection{Setup}
The formalism presented here requires a rotation curve, accretion history, and fixed inner and outer radii to be specified before the simulation is run. Since we employ a logarithmic computational grid, there is little cost to extending the outer radius out to $20$ (as opposed to $10$) kpc. This allows us to follow the transition of the outer disk from somewhat molecular at high redshift to atomic at low redshift. For the inner truncation radius, we take $r_0=0.01 R = 200 pc$. The exact value will affect the quantitative results within a few kpc of the center of the disk, but the exact results of the simulation in this region should be taken with a grain of salt anyway. Here $\sigma_*$ reaches a similar order of magnitude as the circular velocity, which we take to be independent of radius, $v_\phi(r)=220$ km/s, so our treatment of this region as a thin disk is not valid. Moreover, the inner boundary value for the torque equation, which we take to be zero - no torque is exerted by the region within the truncation radius on the disk - could easily be some small but non-zero value.

The accretion history employs the fitting formula from \citet{Bouche:2010xz}, namely
\begin{equation}
\label{eq:cosmoAcc}
\dot{M}(t) = 7\ \epsilon_{in}\ f_{b,0.18}\ M_{h,12}^{1.1}\ (1+z)^{2.2}\ M_\odot/yr
\end{equation}
where $M_{h,12}$ is the halo mass in $10^{12} M_\odot$, $f_{b,0.18}$ is the baryon fraction of the accreting matter normalized to $18\%$, and $\epsilon_{in}$ is zero for $M_{h,12}>1.5$ but varies linearly in time from $0.7$ down to $0.35$ between redshift $2.2$ and $1$. Before redshift $2.2$, $\epsilon_{in}=0.7$, and after redshift $1$, $\epsilon_{in}=0.35$. We choose $f_{b,0.18}=1$, and an initial halo mass which will grow to be about $10^{12} M_\odot$ at redshift zero. The formula governing the growth of the halo mass is given in the same paper,
\begin{equation}
\dot{M}_h = 34.0\ M_{h,12}^{1.14}\ (1+z)^{2.4}\ M_\odot/yr,
\end{equation}
so an initial halo mass of $M_{h,12}=0.27$ at $z=2$ produces a Milky Way-analogue galaxy with $M_{h,12}\approx1$ at $z=0$. \textcolor{black}{We note that some of the baryonic accretion may go into expanding the outer radius of the disk, instead of being transported inward, which would reduce the accretion rate below the estimate given in equation \eqref{eq:cosmoAcc}. However, since the baryonic mass of galactic disks outside 20 kpc is generally a negligible fraction of the total, this clearly cannot be a large effect, and the error we make by neglecting it is small compared to the general uncertainty in the cosmological accretion rate.}

In addition to these functions, there are several free parameters controlling various physical processes in the disk. The star formation efficiency \textcolor{black}{per} freefall time is $\epsilon_{ff}=\textcolor{black}{0.01}$. \textcolor{black}{T}he mass loading factor of winds ejected from the galaxy in proportion to the star formation rate is $\mu=1$\textcolor{black}{, chosen to roughly correspond to observations \citep{Erb:2008mu}}. The fraction of turbulent energy in the gas which will decay in a scale height crossing time is $\eta/1.5 = 1$. The time scale for a $Q_*=Q_{lim}-1$ population to approach $Q_*=Q_{lim}$ is $T_{mig}=2$ local orbital periods, and the value of $Q_*$ below which the stars are subject to transient spiral instabilities is $Q_{lim}=\textcolor{black}{2.5}$. For computational convenience, we use $Q\approx Q_{RW}$ to evaluate the disk's stability. We will explore the sensitivity of the results to these parameter choices in future work. Here our goal is merely to demonstrate the method and its results.

\textcolor{black}{The value of $Q$ everywhere in the disk is fixed to $Q_f$. Theoretically $Q$ is expected to be self-regulated to a value of order unity. Formal stability criteria derived from the perturbed equations of motion for infinitely thin disks find the disks to be unstable when $Q<1$, so the marginal stability which we assume here would imply $Q=1$. However, recent work by \citet{Elmegreen:2011sa} has shown that for a realistically thick disk where the gas cools on the order of a dynamical time, a marginally stable value of $Q$ is closer to 2 or 3. This is consistent with the observational evidence compiled by \citet{Romeo:2011re} for nearby spiral galaxies, namely that when $Q_{RW}$ for these disks is calculated, the values typically fall between 2 and 3 for most galaxies at most radii. Thus we adopt $Q_f=2$ as a fiducial value.}

Finally, to specify the initial conditions fully, one must choose an initial gas fraction and a ratio of stellar to gas velocity dispersion. Since the only way the stellar velocity dispersion can decrease is by mixing it with a lower-velocity dispersion population, it is reasonable to expect this ratio to be greater than unity. The simplified models of gravitationally unstable galaxies evolving from $z \gg 1$ discussed in \citet{Cacciato:2011ab} suggest that by $z\sim 2$, this ratio $\phi_0$ is a few, so we adopt $\phi_0=2$.

\subsection{\textcolor{black}{Disk-Average Quantities}}
Before considering the radial structure of the disk, let us consider the evolution of the galaxy as a whole between $z=2$ and $z=0$. Our model does not allow the rotation curve or outer radius of the disk to evolve in time. However, over this redshift range, the circular velocity (assuming a constant spin parameter) will evolve by less than about $10\%$ \citep[e.g.][]{Cacciato:2011ab}. Meanwhile, the position of the outer edge of the disk has a minimal effect on its evolution, so long as the outer edge of the star-forming disk is resolved. At larger radii than this, there is so little star formation that the gas is free to flow inwards at a constant rate and arrive at the edge of the star-forming disk unaltered by its passage through the HI disk.

\begin{figure}
	\centering
	\includegraphics[width=8.89 cm]{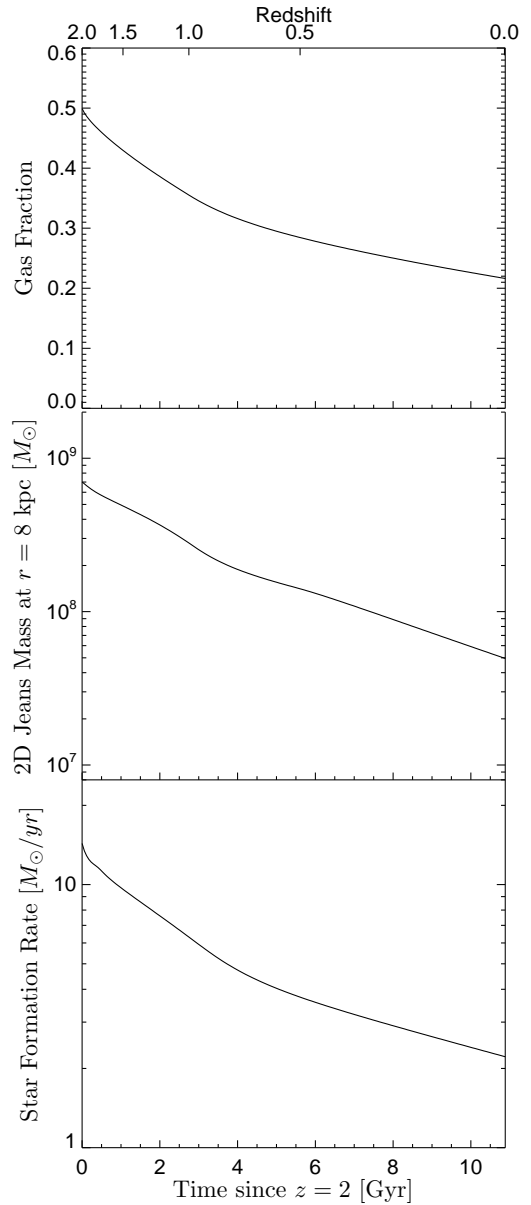}
	\caption{Time evolution from the beginning to the end of the fiducial simulation of the radially-integrated gas fraction, 2D Jeans mass at $r=8$ kpc, and the radially-integrated star formation rate.}
	\label{fig:ts}
\end{figure}

The primary changes in the disk are the steady decline in the accretion rate, and the steady formation of stars. For the fiducial model, $\dot{M}_{ext}(t)$ drops smoothly from about 13 M$_\odot/yr$ at $z=2$ to 2 M$_\odot/yr$ at $z=0$. This falloff is mirrored in the drop in total gas fraction, two-dimensional Jeans mass, and total star formation rate (figure \ref{fig:ts}). The star formation rate in particular has almost the same numerical value as $\dot{M}_{ext}(t)$, starting off slightly higher and converging to the accretion rate. This is a reflection of the fact that the formed stars can only come from gas that started in the simulation or accreted at a later time, and the initial gas reservoir is depleted in about $1$ Gyr.

Stars, once formed, remain in the disk, while the mass of gas in the disk falls with the cosmological accretion rate. This drives a steady decrease in the gas fraction from its initial value, down to $20\%$. Referring to the equilibrium solution for constant gas fraction (equations \ref{eq:equilibrium_sig} and \ref{eq:equilibrium_col}), and noting that $f_g$ has dropped by a factor of a few, while the accretion rate has dropped by a factor of about $6$, we might expect $\sigma$ to decrease by maybe a factor of $2$, while $\Sigma$ might decrease by more than a factor of $3$.

The two-dimensional Jeans mass \citep{Kim:2002jm} is defined by
\begin{equation}
M_{J}=\frac{\sigma^4}{G^2\Sigma}
\end{equation}
Physically this represents the characteristic mass of a clump of gas which collapses under gravitational instability to form a cluster of stars. Its steady decrease with time reflects the cooling of the disk, which allows smaller regions to collapse. This is the phenomenon that explains why $z\sim 2$ galaxies contain giant clumps far larger than the biggest GMCs in present-day Milky Way-like galaxies. As a practical matter, this means that the typical size of star clusters steadily decreases, so, coupled with the fact that a clump of gas can form stars with at most tens of percent effiency, clusters with $M>10^6 M_\odot$ are unable to form in today's quiescent spirals. In the fiducial model, $M_J \sim 2 \cdot 10^7 M_\odot$ at $r=8$ kpc. The decrease in the upper envelope of cluster mass with time is consistent with the arguments made by \citet{Escala:2008cl}.

\begin{figure}
	\centering
	\includegraphics[width=8.89 cm]{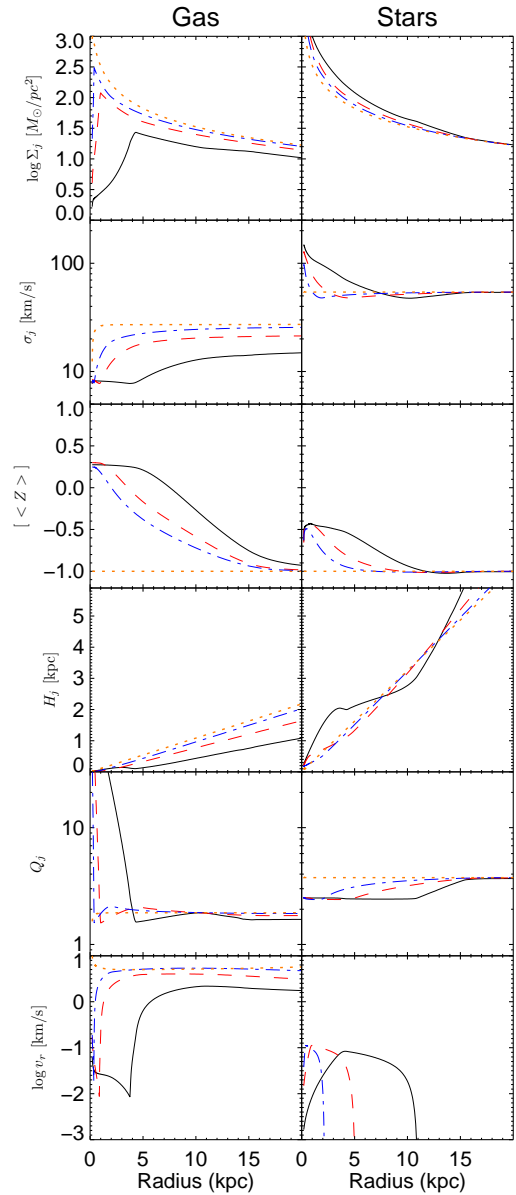}
	\caption{A direct comparison of the gas and stellar components as a function of radius at redshifts 2 (orange dotted), 1.5 (blue dot-dash), 1 (red dashed), and 0 (black solid). The gas cools and depletes, while the stars accumulate and heat. The expanding stabilized region of the disk is evident in the dramatic decrease in gas transport velocity, large $Q_g$, and $\sigma\rightarrow\sigma_t$. The outward movement of the region where stars form and migrate follows the peak in gas column density - $Q_*$ approaches $Q_{lim}=\textcolor{black}{2.5}$, the stellar metallicity gradient steepens, and the stellar scale height flattens. }
	\label{fig:rtsc}
\end{figure}

\subsection{Radial Structure of the Disk}

\begin{figure}[h!]
	\centering
	\includegraphics[width=8.89 cm]{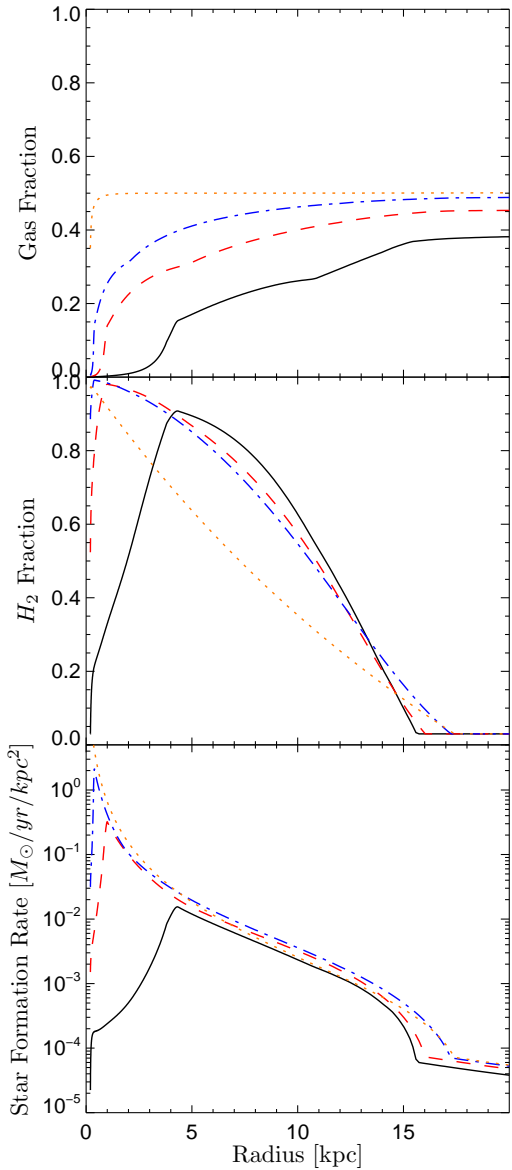}
	\caption{Radial profiles of quantities at redshift 2 (dotted), 1.5 (dot-dashed), 1 (dashed), and 0 (solid). The peak of $f_{H_2}$ and hence the star formation rate move outwards as the simulation evolves, as the gas further in has been depleted and cannot be replenished.}
	\label{fig:rtsg1}
\end{figure}

\begin{figure}[h!]
	\centering
	\includegraphics[width=8.89 cm]{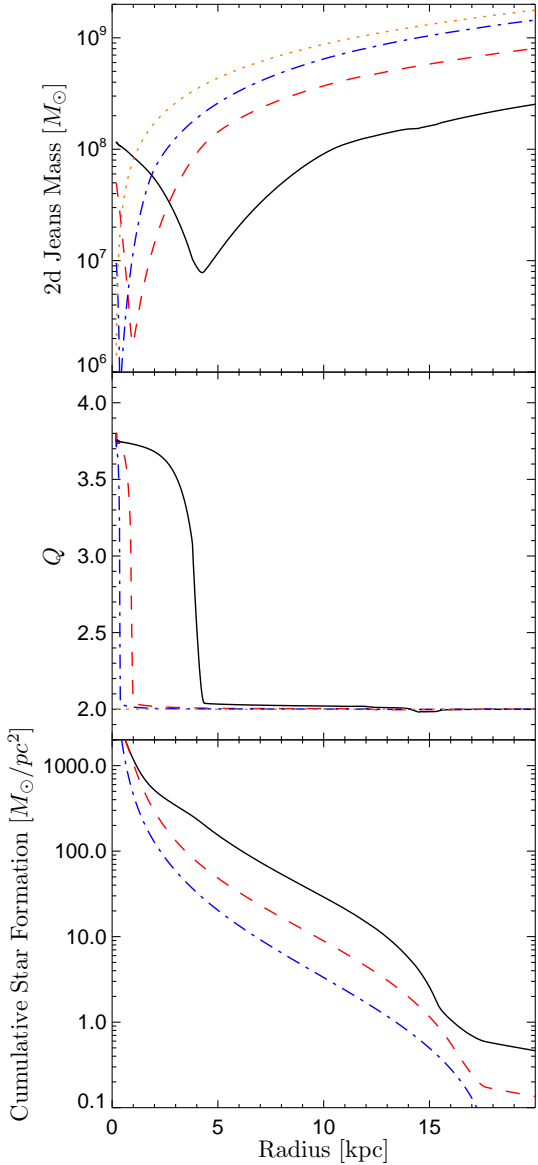}
	\caption{Radial profiles of quantities at redshift 2 (dotted), 1.5 (dot-dashed), 1 (dashed), and 0 (solid). Within the star-forming region, the size of the Jeans mass decreases steadily, but increases at the center of the disk owing to the extremely low gas column densities. The two-component $Q$ value transitions from unity in the gas (both $H_2$ and HI) dominated regions to $Q=Q_{lim}\textcolor{black}{T=15/4}$ in the stellar-dominated component. }
	\label{fig:rtsg2}
\end{figure}

We show the radial structure of our fiducial disk in figures \ref{fig:rtsc}, \ref{fig:rtsg1} and \ref{fig:rtsg2}. We can understand the qualitative behavior shown in these plots by considering the processes that drive the evolution. The two most important drivers are that $Q=1$ almost everywhere at all times, and that stellar migration tends to self-regulate the stars such that $Q_* = Q_{lim}$ - recall that $Q_{lim}$ is a free parameter, below which stars are subject to transient spiral instabilities. If $Q_*>Q_{lim}$, stars will form and drive up $\Sigma_*$, decreasing $Q_*$, while if $Q_*<Q_{lim}$, the stars will migrate inwards increasing $\sigma_*$ and hence $Q_*$. These two restrictions set $Q_g$ to a value somewhat less than $Q_{lim}$, depending on the local ratio $\sigma_g/\sigma_*$. These forces lead the simulations to form three qualitatively distinct regions: a stabilized stellar-dominated region, a star-forming region, and an HI disk.

The radial extent of the star-forming region is more or less set by where the gas is molecular, i.e. $f_{H_2} \approx 1$. This in turn corresponds to where the gas column density is larger than some metallicity-dependent critical value. For our fiducial initial conditions, the disk is molecular out to $r\approx 15$ kpc at $z=2$. Within this radius, almost the entire disk is vigorously star-forming. As time passes, a stellar-dominated central region begins to appear. This occurs because, towards the center of the disk, the gas has short local dynamical times and hence undergoes rapid star formation. In contrast, the inward mass flux of gas required to maintain $Q\approx \textcolor{black}{Q_f}$ is nearly independent of radius. Star formation depletes this gas as it moves inwards, so by the time it reaches the inner region of the disk, not only is there less gas than there would have been neglecting star formation, but it is being consumed faster. In order to maintain a constant $Q$, given that $Q_*\approx Q_{lim}$, the gas must maintain $Q_g$ close to constant. Star formation decreases the gas column density, so to keep $Q_g$ roughly unchanged, the gas velocity dispersion must fall proportionally. Thus the gas velocity dispersion drops fastest in the center of the disk.

By assuming a fixed gas temperature, we essentially set a floor on the value of $\sigma$. When $\sigma$ hits this floor, which happens first at the inner edge of the computational domain (see figure \ref{fig:rtsc}), the radiative loss rate $\mathcal{L}$ approaches zero. The gas no longer loses energy through shocks, and therefore ceases to move inwards. In this situation that region of the disk ceases to become gravitationally unstable, and $Q$ is allowed to rise. Without any means of mass transport, the gas simply depletes as it forms stars. As the gas column density drops off, the stars dominate the local stability of the disk. Since they are constrained to  $Q_* \approx Q_{lim}$ by our assumptions about stellar migration, the overall \textcolor{black}{value of} $Q \textcolor{black}{/T}$ of the disk in this region approaches $Q_{lim}$ as well. 

The third qualitatively distinct region of the disk may be thought of as the HI disk wherein $f_{H_2}$ is low enough that stars form at a relatively slow rate, and gas flows in adhering even more closely to the equilibrium conditions of equations \eqref{eq:equilibrium_sig} and \eqref{eq:equilibrium_col}, which were derived by neglecting star formation in KB10, than in the star-forming region. In essence, the gas is allowed to flow in with a constant mass flux at each radius, since star formation is not depleting the gas significantly. Depending on the initial conditions of the simulation, the column density of stars may be low enough or the velocity dispersion of the stars high enough that $Q_* > Q_{lim}$ for the duration of the simulation. In this situation the overall stability of the disk is almost exclusively determined by the stability of the gas, therefore the gas properties will correspond more closely to the equilibrium values with the gas fraction set to unity.

Looking at the values for $\Sigma$ and $\sigma$ near the solar circle (see figure \ref{fig:rtsc}), we see that they are too high relative to their observed values of approximately $13 M_\odot/pc^2$ and $8$ km/s respectively, though not by more than a factor of two. Moreover, the column density of gas near the center of the disk is lower than observed in the Milky Way. Both of these problems stem from the fact that when $\mathcal{L}\rightarrow 0$, mass transport due to gravitational instability shuts off, whereas the real Milky Way has a number of mechanisms to transport gas into its central regions even when $\sigma\rightarrow \sigma_t$. The gas could be transported by a bar instability from larger radii, or the gas which we assume accretes at the edge of the disk could be accreting directly into the central region of the galaxy. Gas can also be recycled back to the ISM from stars. We assume this occurs instantaneously, so we neglect gas from stars which form farther out in the disk and migrate inwards. Nonetheless, our model qualitatively reproduces the structure of $z=0$ disk galaxies: a central stellar-dominated bulge, an extended star-forming disk, and an outer HI-dominated disk with very little star formation.

\subsection{Stellar Populations}

As the stars form in the fiducial simulation, one can treat them as adding together into a single population for the purposes of evaluating the torque equation, while at the same time evolving a number of passive populations, binned by age, alongside the single population. Only the active population affects the stability of the disk, while the passive populations simply serve as tracers of the stars formed during a particular epoch. This in turn is a reflection of the state of the gas at that time, with the added effect of gradual stellar heating through radial migration.

\begin{figure}
	\centering
	\includegraphics[width=8.89 cm]{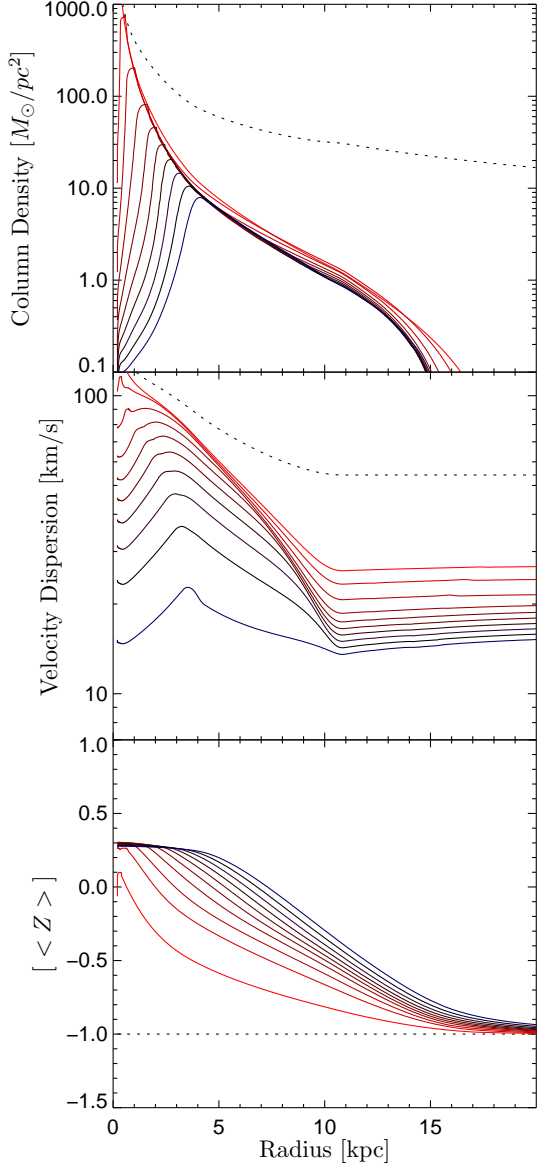}
	\caption{All stellar populations produced in the fiducial model at redshift zero, colored by their age with redder stars older. The ages are linearly spaced in time, so each population is about $1$ Gyr of star formation. The dotted lines represent the initial population of stars, which has only evolved via stellar migration over the whole course of the simulation. Each newer population is less massive, dynamically colder, and has a steeper metallicity gradient than its older analogues.}
	\label{fig:rstpop}
\end{figure}

Stellar migration occurs locally as the result of star formation, since it is star formation which drives $Q_*$ below $Q_{lim}$. It is therefore unsurprising that the stellar populations seem to have very similar column density profiles (see figure \ref{fig:rstpop}) to the star formation rate profile shown in figure \ref{fig:rtsg1}. The primary effect of migration is thus not mass transport inwards, so much as an increase in the velocity dispersion. This can be quite significant - the oldest stars near the center of the disk reach nearly $\sigma_{*,i} = 100$ km/s, which is significantly larger than the gas velocity dispersion at any time in the simulation.

The state of these populations near the solar neighborhood at $z=0$ is of particular interest, since these populations are well-observed and display well-known correlations. The velocity dispersions of stars in the solar neighborhood vary from about $17$ km/s for $1$ Gyr-old stars to $\sim 10$ Gyr-old stars with $\sigma_*\approx 37$ km/s \citep{Nordstrom:2004av,Holmberg:2009gc}. The theoretical explanations for this correlation go back to \citet{Spitzer:1953he} and generally center around the scattering of stars by molecular clouds and spiral structure, which gradually heats the disk. Other explanations have included minor or major mergers \citep[e.g.][]{Dierickx:2010jm,Bekki:2011mm,Qu:2011nm} and popping star clusters \citep{Assmann:2011pc}. All of these explanations are conceptually trying to do the same thing - form a thick disk from a thin disk. However, a gravitationally unstable disk subject to star formation and a decreasing accretion rate will start with a high gas velocity dispersion that will decrease with time. This will also naturally generate an age-velocity dispersion correlation. \textcolor{black}{This is the scenario presented in the simulations of \citet{2009ApJ...707L...1B}, and in the chemodynamical models of \citet{Burkert:1992th}.}

The age-velocity dispersion produced in our fiducial model may be explained as the combination of \textcolor{black}{two} physical effects. \textcolor{black}{First, the gas velocity dispersion decreases with time as the disk cools. This may be understood from the fact that if $Q$  and $Q_*$ are self-regulated to constant values, then $Q_g$ must remain close to constant, and so if $\Sigma$ decreases, so must $\sigma$. As the gas cools, the stars it forms will be cooler than previous generations of stars, leading to an age-velocity dispersion correlation. The second effect is the heating of stars via transient spirals to maintain $Q_*=Q_{lim}$.  Although this is a scattering process which heats stars over time, there is never a thin disk which gradually forms a thick disk.}

\begin{figure}
	\centering
	\includegraphics[width=8.89 cm]{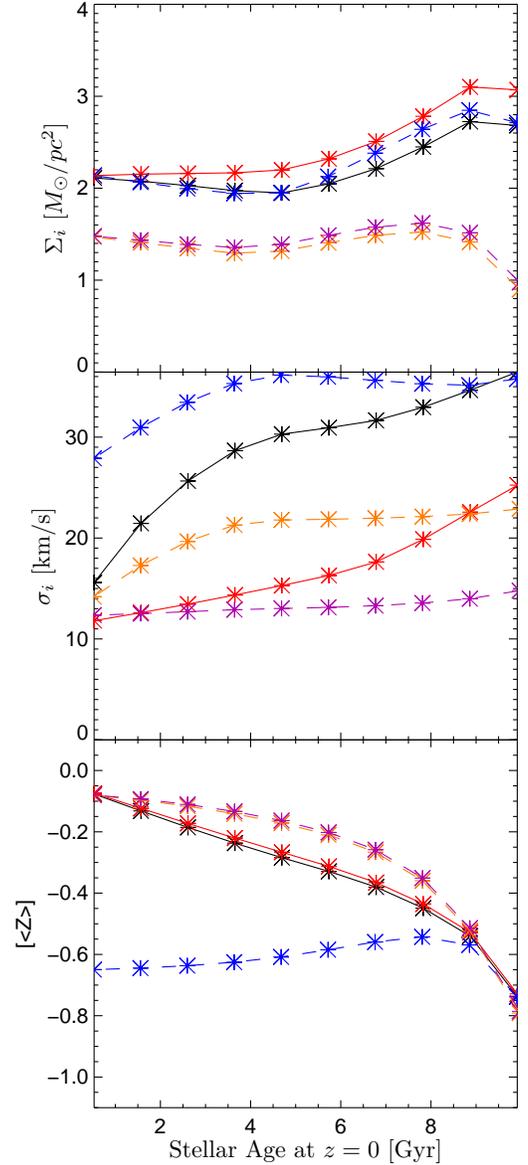}
	\caption{Properties of stellar populations as a function of their age at a radius of 8 kpc at redshift zero. Note that the stars comprising the initial condition of the disk are not plotted here. Each line shows the result of a different model: the fiducial model (black), stellar migration off (red), high constant accretion (orange), low constant accretion (blue), stellar migration off and low constant accretion (purple). The models with constant accretion history are dashed. Every simulation produces an age-velocity dispersion correlation via some combination of increasing $\sigma_{*,i}$ of existing stars or decreasing $\sigma$, which makes the younger stars dynamically cooler.}
	\label{fig:stpop_sol}
\end{figure}
To better discern the importance of each of these effects, we can compare the stars produced by the fiducial model to runs with certain effects artificially turned off. The high and low constant accretion rate models shown in figure \ref{fig:stpop_sol} have $\dot{M}_{ext}(t)= 12.3 M_\odot/yr$ and $\dot{M}_{ext}(t) = 2.34 M_\odot/yr$ respectively, corresponding to the accretion rates at the beginning and end of the fiducial simulation. For simulations where migration is turned off, we plot the properties of the stars at their epoch of formation, rather than their properties at $z=0$. Thus the dynamical effects of migration as it affects the stability of the disk remain unchanged as compared with the fiducial simulation. Figure \ref{fig:stpop_sol} shows explicitly that the age-velocity dispersion correlation is strongly affected by the accretion history and the presence of stellar heating. All of the scenarios are able to generate some age-velocity dispersion correlation.  Even the case with no stellar heating and a constant accretion rate produces one as the result of a fall in $\Sigma$, and hence $\sigma$, as a result of star formation.

\section{Discussion}
Starting from conservation laws and simple assumptions about the gravitational stability of the disk, we have derived evolution equations for the radial profile of a two-component disk. Compared to semi-analytic models, this approach has the advantage that the vast variation in the state variables as a function of radius is resolved rather than averaged over the whole disk. This improvement comes with additional computational costs; however, these are not severe - even using the full Rafikov $Q$ and multiple stellar populations, the code can evolve a disk from $z=2$ to $z=0$ on a single processor in a few days, and using the \citet{Romeo:2011re} approximation to $Q$ reduces the computation time to under an hour. 

This paper is primarily meant to introduce our methodology. However, the fiducial model demonstrates a key point which is often overlooked in galaxy evolution and studies of the thick disk, namely that thick disks need not be formed from thin disks. An age-velocity dispersion correlation appears in our simulation, not because of external perturbers, mergers, or \textcolor{black}{gradual heating of a thin disk}, but because $\sigma$ decreases with time and newly formed stars induce transient instabilities in the disk \citep[see also][]{Burkert:1992th}. Both of these processes are strongly dependent on the cosmological situation in which the disk finds itself, that is, its accretion history. Simulations of isolated thin disks which are gradually heated are therefore unrealistic, in the sense that they are missing the most important drivers of thick disk formation. \textcolor{black}{The smooth increase in stellar velocity dispersion with age produced in our simulations agrees qualitatively with recent observations which demonstrate the lack of a distinctive bimodality between thick and thin disk stars \citep{Bovy:2011rh}.}

This approach has several further applications which we intend to explore in future work. For Milky Way-like galaxies, even modern chemodynamical models with sophisticated treatments of stellar migration and evolution rely on highly parameterized treatments of gas inflow in the disk \citep{Schonrich:2009dg,Spitoni:2011rf}. If the gas evolves to keep the disk marginally gravitationally unstable, its movement in the disk is not this simple - it depends on the evolution of the full non-linear set of equations we have derived here. By accounting for the diffusion of stars in radius as the result of scattering across corotation resonances \citep{Sellwood:2002kk}, our model could be extended to model the Milky Way in detail and compare directly \textcolor{black}{with} observations of the metallicity gradient as a function of height above the disk \citep{Cheng:2011fe}, the age-velocity dispersion correlation \citep{Holmberg:2009gc}, the age-metallicity relation or lack thereof \citep{Edvardsson:1993am},\textcolor{black}{ and the radial and vertical stellar density distributions \citep{Bovy:2011rl}}.

Galaxy bimodality - the separation of galaxies into a blue cloud of star-forming galaxies and a red sequence of ellipticals - is often viewed as an evolutionary sequence. Blue cloud galaxies gradually accrete gas and smaller galaxies, which fuel star formation. Some of these galaxies will undergo major mergers, leaving red and dead elliptical galaxies. These early-type galaxies can subsequently undergo dry mergers, which extend the red sequence to include extremely massive galaxies. Beyond this canonical view, \citet{SanchezAlmeida:2011cl} have noted the existence of a significant population of red spirals. 

By taking more realistic accretion histories from cosmological simulations, we expect that a certain fraction of disks in the course of their lifetimes will experience a period of low accretion during which they will exhaust their gas supply and become redder, only to return to the blue cloud with the resumption of higher accretion rates. \textcolor{black}{} Given a set of realistic non-smooth \textcolor{black}{but quiescent} accretion histories\textcolor{black}{, appropriate for a large fraction of sub-$L_*$ galaxies}, we may therefore be able to reproduce aspects of phenomenology from the local universe out to $z=2$ as semi-analytic models do, but with the added benefit of a physical treatment of the disk dynamics. 

\acknowledgments
We thank M. Cacciato and A. Dekel for stimulating conversations, \textcolor{black}{and the anonymous referee for a thorough and helpful report.} JF is supported by a Graduate Research Fellowship from the National Science Foundation. MRK acknowledges support from the Alfred P. Sloan Foundation, from the NSF through grant CAREER-0955300, and NASA through Astrophysics Theory and Fundamental Physics Grant NNX09AK31G and a Chandra Space Telescope Grant.  AB thanks his colleagues at the astronomy department at UCSC for their hospitality and support.

\appendix

\section{Non-dimensional Equations}
For the purposes of implementing the governing equations in a numerical code, it is useful to non-dimensionalize the equations. To do so is straightforward, and basically amounts to rescaling lengths to the radius of the disk, velocities to the circular velocity, and mass fluxes to the initial accretion rate of gas from the IGM. We can make the following substitutions, following KB10: $r=x R$, $t=T[2\pi R/v_\phi(R)]$, $\mathcal{T} = \tau \dot{M}_{ext,0} v_\phi(R) R $, $\sigma_j = s_j v_\phi(R)$, and $\Sigma_j= S_j \dot{M}_{ext,0}/(v_\phi(R) R)$. Here the subscript $j$ may refer to gas or one of possibly many stellar populations. With these substitutions, the gas evolution equations \eqref{eq:dcoldt} and \eqref{eq:dsigdt} become
\begin{eqnarray}
\frac{\partial S}{\partial T} &=& \frac{(\beta^2 + \beta + x\beta')\tau' - x(\beta+1)\tau''}{(\beta+1)^2 u x^2} - (f_R + \mu)\frac{\partial S_*}{\partial T}^{SF} \\
\frac{\partial s}{\partial T} &=& -\frac{s}{3(\beta+1)S u x} \tau'' + \frac{(\beta+\beta^2+x\beta') s - 5 s' x (\beta+1)}{3(\beta+1)^2S u x^2}\tau' \nonumber \\
& & + \frac{u(\beta-1)}{3 s S x^3}\tau - \textcolor{black}{\frac{2\pi^2}{3} \eta S K_0 \left(1 + \frac{S_*}{S}\frac{s}{s_*} \right)\left(1 - \frac{s_t^2}{s^2}\right)^{3/2} }
\end{eqnarray}
Primes denote partial derivatives with respect to $x$, and as with dimensional quantities, $S$ and $s$ with no subscript refer to properties of the gas. \textcolor{black}{The dimensionless initial accretion rate is
\begin{equation}
K_0 = \frac{G \dot{M}_{ext,0}}{v_\phi(R)^3}.
\end{equation}}The dimensionless thermal gas velocity dispersion is $s_t$.

Employing the same procedure for the evolution equations of each stellar population's column density, we obtain
\begin{eqnarray}
\frac{\partial S_{*,i}}{\partial T} &=& f_R \left({\frac{\partial S_{*,i}}{\partial T}}\right)^{SF} + {\frac{\partial S_{*,i}}{\partial T}}^{Mig}, \\
\frac{\partial S_{*,i}}{\partial T}^{SF} &=&\textcolor{black}{ 8\pi \sqrt{\frac{2}{3}} f_{H_2} \epsilon_{ff} K_0 \frac{S^2}{s} \sqrt{1+\frac{S_*}{S}\frac{s}{s_*}}}, \\ 
\frac{\partial S_{*,i}}{\partial T}^{Mig} &=& -2\pi y \left( S_{*,i} \frac{y'}{y} + S_{*,i}' + \frac{S_{*,i}}{x} \right) 
\end{eqnarray}
where we have explicitly separated the effects of stellar migration and star formation. The dimensionless radial component of the bulk stellar velocity is $y = v_{r*}/v_\phi(R)$.

Similarly, the velocity dispersion evolution equations are
\begin{eqnarray}
\frac{\partial s_{*,i}}{\partial T} &=&  \frac{\partial s_{*,i}}{\partial T}^{SF} + \frac{\partial s_{*,i}}{\partial T}^{Mig},\\
\frac{\partial s_{*,i}}{\partial T}^{SF} &\approx &    f_R \frac{1}{ 2 S_* s_*} (s^2 - s_*^2) \frac{\partial S_*}{\partial T}^{SF}, \\
\frac{\partial s_{*,i}}{\partial T}^{Mig} &= &-2\pi y \left( \frac{(1+\beta) u^2}{3 x s_*} + s_*'  \right)
\end{eqnarray}
The change in velocity dispersion as a result of star formation is only an approximate relation, since it relies on a first order Taylor series expansion of the exact change in $s_{*,i}$, which in turn requires that $S_{*,i} \gg (\partial S_{*,i}/\partial T)^{Mig} dT$. This condition cannot be satisfied when a completely new population of stars is formed as the simulation crosses into a new age bin, at which time $S_{*,i} = 0$. Therefore we use the exact relation,
\begin{equation}
s_{*,i,new} = \sqrt{\frac{(S_{*,i}s_{*,i}^2)_{old} + f_R (dS_{*,i}^{SF})s^2}{S_{*,i,old} + f_R (dS_{*,i}^{SF})}}
\end{equation}
where $dS_{*,i}^{SF} = dT (dS/dT)^{SF}$

Finally we have the equations describing the transport of metals in the gas,
\begin{equation}
\frac{\partial Z}{\partial T} = - \frac{2\pi}{(\beta+1)x S u} \frac{\partial \ln Z}{\partial x}\tau' + \frac{y_M(1-f_R)}{S}\frac{\partial S_*}{\partial T}^{SF}
\end{equation}
and in a stellar population,
\begin{equation}
\frac{\partial Z_{*,i}}{\partial T}^{Mig}= -2\pi y S_{*,i}'.
\end{equation}
The stellar metallicity change owing to the formation of new stars can be computed exactly as
\begin{equation}
Z_{*,i,new} = \frac{(S_{*,i} Z_{*,i})_{old} + f_R(dS_{*,i}^{SF}) Z}{S_{*,i,old} + dS_{*,i}^{SF}}
\end{equation}

These equations, given a torque $\tau$ and a radial stellar velocity $y$, fully describe the evolution of the system. To obtain these two quantities, we imposed conditions on the evolution of $Q$ and $Q_*$ (equations \ref{eq:torque1} and \ref{eq:heating}). In dimensionless form these partial differential equations are
\begin{mathletters}
\begin{eqnarray}
y' + y \left( - \frac{u^2}{s_*^2}\frac{(1+\beta)}{3x} - \frac{s_*'}{s_*} + \frac{S_*'}{S_*} + \frac{1}{x} \right) &=& \frac{\mbox{max}(Q_{lim}-Q_*,0)u}{2\pi x T_{mig} Q_* } \\
g_2\ \tau'' + g_1\ \tau' + g_0\ \tau &=& g_F
\end{eqnarray}
\end{mathletters}
where the coefficients of the dimensionless torque equation are
\begin{mathletters}
\begin{eqnarray}
g_2 &=& - \frac{s}{3 x S u (\beta+1)} \frac{\partial Q}{\partial s} - \frac{1}{(\beta+1) x u}\frac{\partial Q}{\partial S} \\
g_1 &=& \frac{\beta^2 s + s(x\beta' + \beta) - 5(\beta+1)x s'}{3(\beta+1)^2 x^2 u S}\frac{\partial Q}{\partial s} + \frac{\beta(\beta+1) + x\beta'}{(\beta+1)^2 x^2 u}\frac{\partial Q}{\partial S} \\
g_0 &=& \frac{u(\beta-1)}{3 x^3 s S} \frac{\partial Q}{\partial s} \\
g_F &=& \textcolor{black}{\frac{2\pi^2}{3} \eta S K_0 \left(1 + \frac{S_*}{S}\frac{s}{s_*} \right)\left(1 - \frac{s_t^2}{s^2}\right)^{3/2} } \frac{\partial Q}{\partial s} + (f_R + \mu) \frac{\partial S}{\partial T} ^{SF} \frac{\partial Q}{\partial S} \\
\nonumber
& &- \sum_i \left(\frac{\partial S_{*,i}}{\partial T} \frac{\partial Q}{\partial S_{*,i}} + \frac{\partial s_{*,i}}{\partial T}\frac{\partial Q}{\partial s_{*,i}}  \right)
\end{eqnarray}
\end{mathletters}
Both partial differential equations require an outer boundary condition, which essentially specifies the flux of each type of material at the edge of the disk. The mass flux of the gas is specified by some accretion history $\dot{M}_{ext}(t)$,
\begin{equation}
\tau ' (x=1) = - \left(\frac{\dot{M}_{ext}(t)}{\dot{M}_{ext,0}}\right) (1+\beta(x=1)),
\end{equation}
while the flux of stars is set to zero via $y(x=1) = 0$. The torque equation also requires an inner boundary condition, which we take to be $\tau(x=x_0)=0$

\bibliography{libJul3}

\clearpage

\end{document}